\documentclass[12pt]{JHEP3}
\usepackage{graphics}
\usepackage{epsfig}
\usepackage{amsfonts}
\usepackage{psfrag}
\usepackage{amscd}
\newcommand{\be}[1]{ \begin{equation}\label{#1} }
\newcommand{\ee}{\end{equation}}
\newcommand{\bea}[1]{\begin{eqnarray}\label{#1} }
\newcommand{\eea}{\end{eqnarray}}
\newcommand{\eq}[1]{(\ref{#1})}
\def\ZZZ{{\hskip-3pt\hbox{ Z\kern-1.6mm Z}}}
\def\zzz{{\hskip-3pt\hbox{ z\kern-1mm z}}}
\title{
On walls of marginal stability in ${\cal N}=2$  string theories}
\author{Justin R. David\footnote{On lien from Harish-Chandra Research Institute, Allahabad, India.}  \\
Centre for High Energy Physics,\\
Indian Institute of Science, \\
Bangalore 560012, India.\\
\email{justin@cts.iisc.ernet.in}
}

\abstract{ We study the properties of walls
 of marginal stability for BPS decays in a class of ${\cal N}=2$ theories.
These theories arise in ${\cal N}=2$ string compactifications
obtained as freely acting orbifolds of ${\cal N}=4$ theories, such theories include the 
STU model and the FHSV model.  
The cross sections of these walls for a generic decay  in the axion-dilaton plane reduce to 
lines or circles. 
From the continuity  properties of walls of marginal stability 
we show that central charges of BPS states do not vanish in the interior of the moduli space.
Given a charge vector of a  BPS state  corresponding to a large black hole in these theories, we show that 
all walls  of marginal stability intersect at the same point in the lower half of the axion-dilaton plane. 
We isolate a class of decays  whose walls  of marginal stability always
lie in a region bounded by walls formed by decays to small black holes. 
This enables us  to isolate a region in moduli space for which no decays 
occur within this class.
We then study entropy enigma decays for such models 
and  show that for generic values of the moduli, that is when moduli 
are of order one compared to the charges, entropy enigma
decays do not occur in these models.
}

\begin{document}
\section{Introduction}

The spectrum of BPS states in 
a supersymmetric theory are known to jump across walls of marginal stability
when asymptotic moduli are varied
\cite{Seiberg:1994rs,Seiberg:1994aj}.
The jumps in the spectrum across the line of marginal stability is given by the 
wall crossing formula \cite{Denef:2000nb,Denef:2007vg}. 
Any proposal for the BPS spectrum should incorporate these jumps. 
Recent studies have led to a good understanding of  spectrum of $1/4$ BPS dyons in 
a large class of ${\cal N}=4$ string theories \cite{David:2006ud,Banerjee:2008pu, Dabholkar:2008zy}
\footnote{See \cite{Sen:2007qy} for a review.}. All walls of marginal stability 
of $1/4$ BPS dyons with 
co-dimension one have been classified in these theories 
\cite{Sen:2007vb, Mukherjee:2007nc}
It has also been shown that
 jumps in the spectrum of  $1/4$ BPS  states across these walls are 
consistent with the wall crossing formula
 \cite{Sen:2007vb, Dabholkar:2007vk,Cheng:2007ch, Sen:2007pg,Sen:2008ht}

For BPS states in 
 ${\cal N}=2$ theories a similar understanding has yet to emerge.
In \cite{David:2007tq} a proposal for the spectrum of class 
of $1/2$ BPS state in the STU model 
was put forward. The  first subleading corrections in entropy for
large charges evaluated from this proposal agrees with that 
evaluated using the Hawking-Bekenstein-Wald formula 
including the Gauss-Bonnet term. 
For large charges the partition function 
proposed for the STU model   also reduces to 
  the OSV form \cite{Ooguri:2004zv} 
   on performing the Laplace transform with respect to the 
  electric charges \cite{Cardoso:2008ej}. The pre-factor which arises in this
  Laplace transform agrees with that proposed by \cite{Cardoso:2008fr}. 
 In \cite{Cardoso:2008ej}, the  proposed partition function for the 
 STU model was argued to be  to be valid 
only for single centered black holes.  Thus the spectrum obtained
from the proposed partition function is valid only when the asymptotic  moduli equals the
moduli at the attractor point. 
It is only for these values of the asymptotic  moduli the single centered black hole is stable and multi-centered configurations do not exist. 
To understand how to extend the partition function to all regions in asymptotic moduli
space it is necessary to study  the walls of marginal stability in this model
and the domains formed by the intersection of these walls. 

To classify all the possible walls of marginal stability
and study the domains formed by them for an arbitrary 
${\cal N}=2$ theory  is in general  a difficult task. This is because
in ${\cal N}=2$ theories all BPS decays are co-dimension one surfaces
while in  ${\cal N}=4$ theories  only decays to small black holes 
are co-dimension one surfaces \cite{Sen:2007vb,Mukherjee:2007nc,Mukhi:2008ry}. In fact unless the surface of marginal stability is a co-dimension one surface, the index which counts the BPS state does not jump \cite{Dabholkar:2009dq}.

 In this paper  given a BPS state 
 specified by a primitive charge vector $(Q, P)$ corresponding to a large black hole
 we  study various properties of walls of marginal
 stability in a class of ${\cal N}=2$ theories. The properties we find will enable
 us to determine the conditions on the charges  of the decay
  and the moduli such that we obtain simple 
domains in moduli space where states are stable under a class of decays. 
The class of ${\cal N}=2$ models we will focus on in this paper are those  theories constructed as freely acting orbifolds of ${\cal N}=4$ theories.  The vector multiplet 
moduli space of such theories is known exactly and is of the form
\begin{equation}
\label{coset}
 {\cal M}_V = \frac{SU(1,1)}{U(1)} \times \frac{SO(2, n)}{ SO(2) \times SO(n)}.
\end{equation}
The axion-dilaton moduli $\tau$, parametrizes the coset $SU(1,1) /U(1)$, while the 
rest of the vector multiplets parametrizes the coset $SO(2, n)/ (SO(2) \times SO(n) )$. 
The  STU model of \cite{Sen:1995ff,Gregori:1999ns} has $n=2$, while the FHSV 
\cite{Ferrara:1995yx}  model has $n=10$. 
Other  models belonging to this class  with $n=4,  6$ have been constructed 
and studied in \cite{Gregori:1998fz}.

There is a  convenient parametrization of the  coset in \eq{coset} which enables us  
write down a simple mass formula for $1/2$ BPS states in these models.  
From this mass formula we show that if the walls of marginal stability is seen
as sections in the $\tau$-plane they will be either circles or lines. 
Using the continuity properties of the walls of marginal stability we show that the
central charges of larges black holes do not vanish in the interior of the moduli space.
Since  the walls are circles or lines in the $\tau$ plane it is  
sufficiently easy to study the intersection of these and find domains bounded by these walls.  
To isolate this class of decays we obtain 
certain properties of the walls of marginal stability  which are 
true for all decays of a given charge vector in these theories. 
Here are some of the  general properties of walls of marginal stability 
of a given charge vector $(Q, P)$ we find in this paper.
\begin{enumerate}
\item{All walls of marginal stability are circles or lines in the $\tau$ plane. }
 \item All walls  of marginal stability 
 meet at the same point in the lower half $\tau$ plane.
\item Walls of marginal stability corresponding to decays to two 
small black holes for any generic moduli always exist. 
\item  To specify a wall uniquely, it is just necessary to provide the 
two points $r_-$ and $r_+$ at which it intersects the real axis in the 
$\tau$ plane. 
\item   The  necessary and sufficient conditions  that  walls 
intersect each other only on the real axis and never in the interior of the 
upper half $\tau$ plane is 
$$
r_+=  \frac{p}{q} , \quad r_- = \frac{p'}{q'}, \quad\hbox{with} \quad pq' -p'q =1,
$$
where $p, q, p', q' \in \ZZZ$. If one of these points is at infinity, then the wall
is   line and it must pass through an integer on the real axis. 
 In fact we show if the  above conditions are not true, 
there is always a wall corresponding to a small black hole decay which
intersects the circle joining $r_-$ and $r_+$ or the lines passing through
 the integer points. 
\end{enumerate}

\begin{figure}
\begin{center}
\psfrag{T1}{$\tau_1$}
\psfrag{T2}{$\tau_2$}
 \epsfig{file=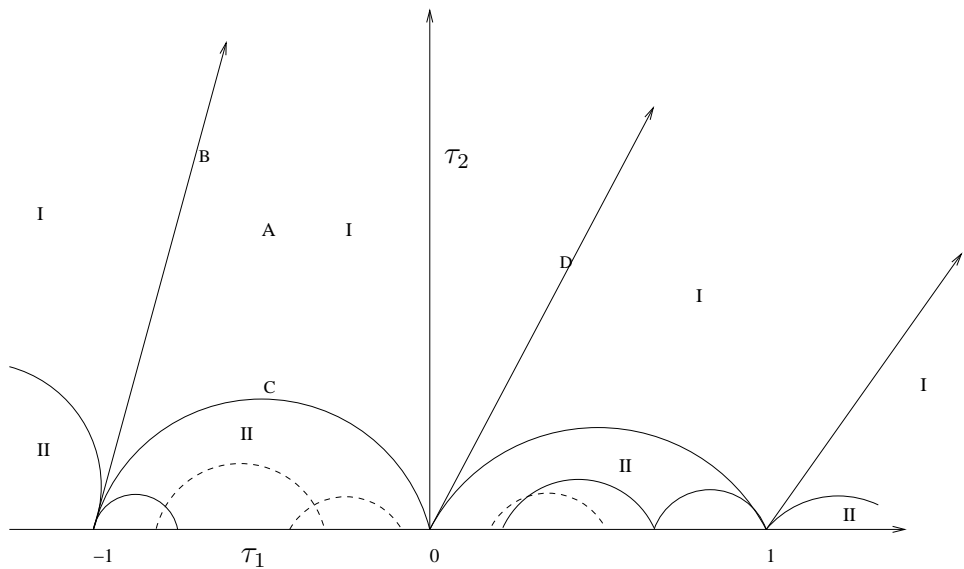,width=5.0in}
\caption{Walls of marginal stability in the upper half $\tau$ plane.
Bold lines correspond to walls coincident with small black hole decays. 
Dashed line correspond to generic decays.}\label{fig-domains}
\end{center}
\end{figure}

Let us now look at the structure of domains formed by  walls of marginal stability. 
The structure of domains plays a role in determining or testing the BPS spectrum. 
If there is domain with no walls of marginal stability, then the BPS spectrum remains
the same in that domain and does not jump. The boundaries of the domain will determine
how the BPS spectrum jumps when one crosses the walls which determine the domain. 
An example of a  simple  domain bounded by walls of marginal 
stability is that found  in \cite{Sen:2007vb} for the case of 
${\cal N}=4$ decays. This domain  also exists  for the ${\cal N}=2$ models under 
consideration. If one restricts the attention to decays to two small black holes,
 the walls  of marginal stability are circles
or lines in the upper half $\tau$ plane.  
Examples of these are drawn in figure  \ref{fig-domains}. The walls which 
correspond to decay to two small black holes are shown with bold lines. 
They intersect  each other only on the real line and at 
at rational points. Consider the region A bounded by the line
joining B and point $-1$ on the real line, the circle joining $-1$ and the origin
and the line passing through the origin and $D$.  In this region there 
are no decays to two small black holes. Similar domains exist
in each interval $[n, n+1], n\in \ZZZ$, for example the domain 
above the circle joining the origin and the point $1$ on the real axis and so on.
In this paper we isolate a class of decays, which include decays to large 
black holes whose walls  of marginal stability always lies in the region II, 
that is in the region bounded by the  small black hole decay corresponding
to circles joining the points $(n, 0)$ and $( n+1, 0)$ on the real line
\footnote{The physical $\tau$ plane is determined by condition $\tau_2\geq 0$.}.
These lines of marginal stability are represented by the dashed lines in 
figure \ref{fig-domains}. 
They can intersect each other or that of walls 
corresponding to  small black decays in the interior of the upper half 
plane at the most once in region II. 
Thus restricting our attention to  these class of decays, 
which include decays to large black holes we see that the region 
 region A continues to be a domain  where there are no decays 
occur among this class. 

As we have a convenient mass formula for the
$1/2$ BPS states in  these class of ${\cal N}=2$ models, 
we can use it to study the entropy enigma found in \cite{Bates:2003vx,Denef:2007vg}.
The enigma results when a given $1/2$ BPS states decays to products whose
entropy is parametrically larger than that of  the initial state.
  By analyzing the various cases we show that such decays are 
not allowed at generic regions of moduli space.  By generic we mean when 
moduli are not scaled parmetrically and  is of order one. This conclusion is consistent with that found in \cite{Denef:2007vg,Denef:2007yn} 
for specific examples, but here we demonstrate it in general for these class of 
${\cal N}=2$ models. 

The organization of the paper is as follows: In section 2 we present the mass formula for $1/2$ BPS states
in the class of ${\cal N}=2$ theories we will be dealing with and show that 
for BPS states, the mass or the central charge does not vanish in the interior of moduli space. 
We then present the conditions for the existence of walls of marginal stability and determine 
their equations. 
In section 3. 
we first  show that all decays  of a given charge vector meet at a  common point in the lower half
$\tau$ plane ($\tau_2<0$) . 
We then study small black hole decays of a given charge vector and show that walls
corresponding to these decays always exist. We  enumerate the conditions necessary
for the existence of a wall corresponding to a generic decay.
We find the necessary and sufficient conditions such two walls never meet in the 
interior of the upper half $\tau$ plane.  
We then  determine the conditions which  isolate a class of decays
whose walls lie in region  below that bounded by small black hole decays.
In section 4 we study entropy enigma decays using the simple mass 
formula for these models  and show that at generic regions of moduli 
these models do not admit such decays.  The Appendix contains some useful results from 
number theory necessary for our purposes.

\section{BPS mass formula and marginal stability}

In this paper we will be dealing with a class of ${\cal N}=2$ theories which are obtained
by freely acting orbifolds of ${\cal N}=4$ theories. 
For these class of ${\cal N}=2$ theories, it
 is instructive to derive the mass formula for BPS states 
from the  ${\cal N}=4$ mass formula. 
Since these theories are constructed from parent ${\cal N}=4$ models by a freely acting orbifold, 
the $1/4$ BPS states as well as the $1/2$ BPS states of the parent theory will be 
BPS states in the orbifolded ${\cal N}=2$ model. Therefore we can obtain the 
BPS mass formula 
in these models using the BPS mass formula of the ${\cal N}=4$ theory.
Given a charge vector $( Q, P)$ the BPS mass formula for ${\cal N}=4$ theories are given by
\cite{Cvetic:1995uj,Duff:1995sm}
\begin{eqnarray}
 \label{massform}
m( Q,  P)^2 &=& \frac{1}{\tau_2} ( Q- \bar\tau P)^T( M + L ) ( Q- \tau P)  \\ \nonumber
& & + 2 \left[ ( Q^T(M +L)Q)( P^T(M+L) P) - ( P^T(M+L) Q)^2 \right]^{1/2}.
\end{eqnarray}
Note that in this case $Q, P$ is a charge vector belonging to a Narain lattice with 
signature $\{ (-1)^r, (1)^6 \}$ \footnote{For Heterotic on $T^6$, $r= 22$}. 
In \eq{massform}
 we have we have to choose the  branch such that the square root is positive. This guarantees 
that the mass formula is given by the larger of the two central charges of the ${\cal N}=4$ model. 
$M$ refers to the asymptotic moduli of   the vector multiplets of the ${\cal N}=4$ theory and 
$\tau$ refers to the  asymptotic dilaton-axion moduli. 
For the class of  ${\cal N}=2$ models under consideration in this paper  the vector multiplet moduli space is
of the form 
\begin{eqnarray}
 \label{vec-moduli}
{\cal M}_V &=&  {\cal M}_S \times {\cal M}_T, \\ \nonumber
&=& \frac{SU(1,1)}{U(1)} \times \frac{SO(2, n)}{ SO(2) \times SO(n)}.
\end{eqnarray}
Thus to apply the mass formula
in \eq{massform} to these ${\cal N}=2$ models
we just have to restrict  the matrix $M$ so that it  
parametrizes the coset $SO(2, n)/(SO(2) \times SO(n)$ 
Therefore $M$
is a $(2+n) \times (2+n)$ matrix  which satisfies the conditions
\begin{equation}
\label{m-prop}
M^T = M,  \quad M^TL M =L,
\end{equation}
where $L$ is the diagonal$(2+n) \times (2+n)$ matrix given by
$$
L = {\rm{ Dia}} ( -1^n,     1^2).
$$
The charge vectors for the ${\cal N}=2$ models take values in a Narain lattice
with the same signature as $L$. 
We now proceed to give an explicit parametrization of $M$ as in \cite{David:2007tq}. 
First introduce
$n+2$ complex numbers $w_I$  satisfying the constraint
\begin{equation}
\label{constraint}
 - \sum_{I=1}^n w_I^2 + w_{n+1}^2 + w_{n+2}^2 =0,
\end{equation}
together with the identification $w_I \sim c w_I$, where $c$ is a complex number. 
Note that the constraint in \eq{constraint} and the identifications of $w$'s up to complex scalings
reduce the number of independent parameters to $2n$ which is the required number of 
variables to parametrize the moduli space ${\cal M}_T$. 
Using the scaling degree of freedom, the constraints in \eq{constraint} can be 
solved by introducing $n$ complex numbers $(y^+, y^-,  \vec  y)$ where $\vec y$ is a $n$ dimensional
vector. These variables are related to the $w_I$'s by 
\begin{eqnarray}
 \label{w-y-relation}
w_I = y_I, \;\; I = 1, \cdots n-2, \qquad  w_{n-1} = \frac{1}{\sqrt{2}} ( y^+ - y^-), \\ \nonumber
w_{n} = 1 + \frac{y^2}{4}, \qquad w_{n+1} = \frac{1}{\sqrt{2}} ( y^+ + y^-), \\ \nonumber
w_{n+2} = -1 + \frac{y^2}{4}, \qquad y^2 = 2 y^+y^- + \vec y^2.
\end{eqnarray}
 It is easy to see that these values of $w_I$ satisfy the constraint in \eq{constraint}. 
The above parametrization amounts to scaling $w_n - w_{n+2}$ such that its value is 
fixed to be $2$. Using the above solution of the constraint \eq{constraint} it can be 
seen that 
\begin{eqnarray}
 \label{cond-1}
& & - \sum_{I=1}^n |w_I|^2 + |w_{n+1}|^2 + | w_{n+2}|^2 = 2 Y, \\ \nonumber
& & {\rm where} \quad  Y = ({ \rm{Im}} y  )^2 = 2 y_2^+ y_2^- - \vec y_2^2. 
\end{eqnarray}
Here the subscripts $2$ in $y$ refer to its imaginary part. $Y$ is related to the K\"{a}hler potential 
on the moduli space by
\begin{eqnarray}
 K = - \log Y.
\end{eqnarray}
We can now parametrize the  matrix $M$ in \eq{m-prop} as follows
\begin{eqnarray}
 M = L \tilde M L - L,  \qquad 
\tilde M_{IJ}  = \frac{ w_I \bar w_J +\bar w _I w_J}{Y}.
\end{eqnarray}
Using the above parametrization of $M$ one can easily demonstrate its properties 
\eq{m-prop} using \eq{constraint} and \eq{cond-1}. 

We use the above parametrization of the asymptotic
moduli matrix  $M$ in terms of $w$'s in the mass formula  \eq{massform}. 
It is first instructive to see what the terms in the square root in \eq{massform} reduces to 
\bea{subman1}
& & 2 \left[ ( Q^T(M +L)Q)( P^T(M+L) P) - ( P^T(M+L) Q)^2 \right]^{1/2} \\ \nonumber
&=& 2 \left[ \frac{4 |Q\cdot w|^2 |P\cdot w|^2 }{ Y^2} - 
\frac{\{ ( Q\cdot w) ( P\cdot \bar w) + ( Q\cdot \bar w) ( P\cdot w)\}^2 }{Y^2}
\right]^{1/2}, \\ \nonumber
&=& \frac{2}{Y} \left[ (-) \left\{ ( Q\cdot w ) ( P\cdot \bar w)  - ( Q\cdot \bar w)(P\cdot w)\right\}^2 \right]^{1/2}, \\ \nonumber
&=&\pm \frac{  2i}{Y } 
\left[ ( Q\cdot w)( P\cdot \bar w) - ( Q\cdot \bar w) ( P\cdot w) \right], \\ \nonumber
&=& \mp  \frac{  4}{Y } {\rm{ Im} } ( ( Q\cdot w) (P\cdot \bar w ) ).
\eea
Since we have to choose the positive square root we see that for 
${ \rm{ Im} } ( ( Q\cdot w) (P\cdot \bar w ) ) <0$
we have to choose the $-$ ve sign in the last line of \eq{subman1} and 
for ${ \rm{ Im} } ( ( Q\cdot w) (P\cdot \bar w ) ) >0$ we have to choose the $+$ ve sign in 
the last line of \eq{subman1}.  We summarize this in the equation below
\begin{eqnarray}
 \label{cases}  
&  & \left[ ( Q^T(M +L)Q)( P^T(M+L) P) - ( P^T(M+L) Q)^2 \right]^{1/2},
\\ \nonumber
 & & =  \frac{  4}{Y}  \left| {\rm{ Im}}  ( ( Q\cdot w) (P\cdot \bar w ) ) \right|.
\end{eqnarray}
Let us now proceed and derive the BPS mass formula 
The  first term in the mass formula \eq{massform} can be written as 
\begin{eqnarray}
\label{subman2}
 & & \frac{1}{\tau_2} ( Q- \bar\tau P)^T( M + L ) ( Q- \tau P) \\ \nonumber
& & =  \frac{1}{\tau_2 Y} \left[
 ( (Q -\bar\tau P) \cdot w )(  ( Q- \tau P)\cdot \bar w) 
 +( (Q -\bar\tau P) \cdot \bar w )(  ( Q- \tau P)\cdot  w) \right].
 \end{eqnarray}
 Now adding \eq{subman1} and \eq{subman2} we obtain
 \begin{eqnarray}
\label{finalform1}
 m( Q,  P)^2 &=& \frac{2}{\tau_2Y} | (Q -\tau P)\cdot w|^2 , \qquad {\hbox{for}} 
 \;\; {\rm{ Im}}   ( ( Q\cdot w) (P\cdot \bar w ) ) <0, \\ 
\label{finalform2}
 m( Q,  P)^2&=& \frac{2}{\tau_2 Y} |( Q -\bar\tau P)\cdot w|^2 , \qquad{\hbox{for}}
 \;\;{\rm{ Im} }  ( ( Q\cdot w) (P\cdot \bar w ) ) >0.
 \end{eqnarray}
 The first equality arises on choice of the $-$ sign in \eq{subman1}, while
 the second equality arises on the choice of $+$ sign in \eq{subman1}. 
For convenience let us  define 
\begin{equation}
 Z(Q, P) =  (Q -\tau P)\cdot w.
\end{equation}
With this definition the BPS mass formula for the first case in \eq{finalform1} can be written as
\begin{equation}
\label{bpsmass}
 m^2(Q, P) = |\tilde Z( Q, P)|^2 = \frac{2}{\tau_2 Y} | Z(Q, P)|^2.
\end{equation}
The above mass formula agrees with that found in \cite{Ceresole:1994cx} 
for ${\cal N}=2$ models with the vector multiplet moduli of the form 
given in \eq{vec-moduli} \footnote{See equation 5.28 of \cite{Ceresole:1994cx}.}

In a given ${\cal N}=2$ theory the mass of BPS states are given  once and for all by one 
formula which is proportional to the absolute value of the central charge. Therefore we
must choose one branch out of the two possible branches given in \eq{finalform1}, \eq{finalform2}.  In this 
paper we will work with the first branch in \eq{finalform1}, this implies that 
 the condition ${\rm{ Im} } ( ( Q\cdot w) (P\cdot \bar w ) ) <0$ must be true through out 
the moduli space for a given charge vector $(Q, P)$. 
In section 2.2 we will  demonstrate that once we choose to describe BPS states with the 
branch \eq{finalform1}, then the 
condition   ${\rm{ Im} } ( ( Q\cdot w) (P\cdot \bar w ) ) <0$ remains true for BPS states
for all asymptotic moduli $w$. 

\subsection{Marginal stability}

In this section we state  the conditions for marginal stability of a given BPS state and 
show that  the co-dimension one surfaces seen 
as sections  for fixed  $w$ moduli 
in the $\tau$-plane are either circles or lines. 
Consider the marginal  decay  of a BPS primitive charge vector  corresponding to a 
large black hole  with 
$Q^2 >0, P^2>0$  and $Q^2 P^2 -(Q\cdot P)^2 >0$  \footnote{In this paper we will 
be dealing only with primitive charge vectors, that is $(Q, P)$ which cannot be 
written as an integral multiple of another vector.}given by 
\begin{equation}
 \left(\begin{array}{c}
 Q \\   P 
\end{array}
\right) = 
\left(\begin{array}{c}
 Q_1 \\  P_1 
\end{array}
\right)
+ 
\left(\begin{array}{c}
 Q_2 \\  P_2
\end{array}
\right).
\end{equation}
If the above decay is marginally allowed then 
the sum of the mass of the products equals the mass of the initial
state, therefore we have the equation
\begin{equation}
 m(  Q,  P) = m( Q_1,  P_1) + m(  Q_2,  P_2) 
\end{equation}
Examining  the mass formula in \eq{bpsmass}, we see that 
this implies the complex numbers $$Z(Q, P),\quad Z(Q_1, P_1), \quad Z(Q_2, P_2),$$  are all co-linear. 
This leads to the following two conditions
\begin{eqnarray}
\label{imcond}
 { \rm Im} (Z_1 \bar Z_2) = 0, \\ 
\label{recond}
{\rm Re } ( Z_1 \bar Z_2) >0,
\end{eqnarray}
where
\begin{eqnarray}
 Z_1 = ( Q_1 - \tau P_1) \cdot w, \qquad Z_2 = ( Q_2 - \tau P_2) \cdot w, \qquad
Z = (Q- \tau P)\cdot w.
\end{eqnarray}
These two conditions are equivalent to the conditions that phases of the central charges align. 
The first equality imposes the condition that the phases are equal up to a multiple of $\pi$,
this implies that imposing the first equality ensures that the phases can be aligned or off by $\pi$. 
While the second inequality  assures that the phases align. 
Note that the first condition \eq{imcond} can also be written  equivalently as 
\begin{eqnarray}
\label{equivmargin}
 {\rm Im} ( Z \bar Z_1) =0, \qquad \hbox{or}\quad {\rm{Im}} (  Z\bar Z_2 ) =0.
\end{eqnarray}

Let us now proceed to obtain the explicit  equations in the 
$\tau$ plane for a given fixed $w$ moduli. 
Substituting the definitions of $Z$ in the \eq{imcond} we obtain the following equation
\begin{eqnarray}
 \label{circleeq-1}
& & \tau\bar\tau  {\rm Im} [ P_1\cdot w P_2 \cdot \bar w] - 
\tau_1  {\rm Im} [ P_1 \cdot w  Q_2 \cdot \bar w  + Q_1\cdot w P_2 \cdot \bar w]   \\ \nonumber
& & -\tau_2 {\rm Re} [P_1 \cdot w Q_2 \cdot \bar w  -  Q_1\cdot w P_2 \cdot \bar w] 
+ {\rm Im} [ Q_1\cdot w  Q_2\cdot \bar w ] =0 
\end{eqnarray}
In general the above equation \eq{circleeq-1}, is an  equation of a circle in the $\tau$-plane by 
completing the squares. To show this  let us define 
\begin{eqnarray}
 A = {\rm Im} [ P_1\cdot w P_2 \cdot \bar w],  & \;& 
B =  {\rm Im} [ P_1 \cdot w Q_2 \cdot \bar w  + Q_1\cdot w P_2\cdot \bar w] , \nonumber \\
C = {\rm Re} [P_1 \cdot w  Q_2 \cdot \bar w  - Q_1\cdot w P_2 \cdot \bar w], &\;&
D = {\rm Im} [ Q_1\cdot w  Q_2\cdot \bar w].
\end{eqnarray}
Note that in these coefficients are unchanged when one replace $ Q_1 \rightarrow Q, 
P_1\rightarrow P$ or $Q_2 \rightarrow Q, P_2\rightarrow P$ due to the property
that the curve ${\rm{Im}}( Z\bar Z_1)$ can be written as \eq{equivmargin}.  
Completing the squares  in \eq{circleeq-1} one obtains
\begin{equation}
 \left( \tau_1 - \frac{B}{2A} \right)^2 + 
\left( \tau_2 - \frac{C}{2A}\right)^2 = \frac{ B^2 + C^2 - 4AD}{4 A^2}.
\end{equation}
It can be shown using simple algebraic manipulations that
\begin{equation}
 B^2 + C^2 - 4AD = | P_1\cdot w Q_2\cdot w - P_2 \cdot w Q_1 \cdot w|^2.
\end{equation}
Thus for $A\neq 0$ the equation \eq{circleeq-1} is that of a circle, if $A=0$ it reduces to that
of the straight line \footnote{It can be shown that if one examines the 
equation ${\rm{ Im}}(Z_1\bar Z_2)$ for constant $\tau, y^-, \vec y$ moduli in the 
$y^+$ plane, the resultant curve is also a circle. Similarly the  curve is also a 
circle in the $y^-$ plane for constant $\tau, y^+, \vec y$. }.
Let us now examine the second condition for marginal stability given in \eq{recond}. Again 
substituting  the definition of $Z$ into this condition we obtain the equation 
\begin{eqnarray}
 \label{circleeq-2}
& & \tau\bar\tau {\rm Re} [ P_1\cdot w P_2 \cdot \bar w]
- \tau_1 {\rm Re} [ P_1 \cdot w  Q_2 \cdot \bar w  + Q_1\cdot w P_2 \cdot \bar w] \\ \nonumber 
& & + \tau_2 {\rm Im} [P_1 \cdot w  Q_2 \cdot \bar w  - Q_1\cdot w P_2 \cdot \bar w]
+ {\rm Re} [ Q_1\cdot w  Q_2\cdot \bar w]  >0.
\end{eqnarray}
The curve determining the above inequality  is also an equation of a circle, 
this can be seen by completing the squares
\footnote{The inequality ${\rm{Re}} ( Z_1\bar Z_2) >0$
in the $y_+$ plane for constant $\tau, y^-, \vec y$ moduli is also 
determined by a circle. The same statement can be made when the inequality is 
seen in the $y^-$ plane for constant $\tau, y^+, \vec y$. }.
Let us define the following 
\begin{eqnarray}
 A'  = {\rm Re} [ P_1\cdot w P_2 \cdot \bar w], &\; & 
B' = {\rm Re } [ P_1 \cdot w  Q_2 \cdot \bar w  + Q_1\cdot w P_2\cdot \bar w],  \nonumber \\ 
C' = {\rm Im} [P_1 \cdot w  Q_2 \cdot \bar w  - Q_1\cdot w P_2 \cdot \bar w ], &\;&
D' = {\rm Re} [ Q_1\cdot w  Q_2\cdot \bar w ].
\end{eqnarray}
In terms of these variables, the inequality reduces to 
\begin{eqnarray}
A'\left[ \left( \tau_1 - \frac{B'}{2A'}^2 \right) + \left( \tau_2 + \frac{C'}{2A'} \right)^2 \right]
- \frac{| P_1\cdot w Q_2\cdot w - P_2 \cdot w Q_1 \cdot w|^2 }{4A'} >0.
\end{eqnarray}
Here we have used the equality 
\begin{equation}
(B')^2 + (C')^2 - 4A'D'  =  | P_1\cdot w Q_2\cdot w - P_2 \cdot w Q_1 \cdot w|^2.
\end{equation}
Therefore the wall of marginal stability is given by 
the part of the circle in \eq{circleeq-1} which satisfies the inequality in \eq{circleeq-2} and which lies in the 
physical part of the $\tau$ plane.  We call such a wall of marginal stability as a the 
wall corresponding to a  physical decay. 
Let us determine  the point of intersection of the circle which determines the inequality in \eq{circleeq-2} and
the circle in \eq{circleeq-1}. 
From \eq{imcond} and \eq{recond}, we see that they 
intersect at the two points  
\begin{equation}
\label{intersect}
 Z_1 =0, \quad \hbox{or}\quad Z_2 =0.
\end{equation}
We will show in section 2.2  that these points of intersection  never lie in the interior of the
physical moduli space, that is in the interior of the upper half $\tau$ plane $(\tau_2>0)$. 

Let us now compare the conditions for walls of marginal stability with that for the existence of the 
two centered black hole solution with charges $(Q_1, P_1)$, $(Q_2, P_2)$. 
The integrability condition \cite{Denef:2000nb} for the existence of the two centred solution results in the 
following equation 
\begin{equation}
 \label{distance}
R =  \frac{ ( Q_1 \cdot P_2 - Q_2 \cdot P_1)  |\tilde Z (Q,  P)| }{ {\rm{Im}}( Z_1 \bar Z_2) }.
\end{equation}
where $R$ is the distance between the two centers. 
Therefore we see that generically, when $( Q_1 \cdot P_2 - Q_2 \cdot P_1)  \neq 0$
the two centred solutions are stable  and exist at one side of the line ${\rm{Im}}( Z_1 \bar Z_2) =0$
and are unstable on the other side.  At the wall ${\rm{Im}}( Z_1 \bar Z_2) =0$ the distance
between the two centers of the  black holes goes to infinity.

\subsection{Non-vanishing of central charges and ${\rm{ Im}} Q\cdot w P\cdot \bar w  <0$}

It has been argued in \cite{Seiberg:1994rs,Strominger:1995cz,Moore:1998zu}
 that the central charge of a BPS state  corresponding to 
a large black hole never vanishes in  the interior
of the physical moduli space. In fact, it has been shown that  the minimum value of the  modulus 
 of the central charge of a given BPS state occurs  at the attractor value of the moduli
which is proportional to the classical entropy of the corresponding black hole \cite{Moore:1998zu}.
The fact that the central charge of a BPS state never vanishes
at the interior of the physical moduli space  was used recently in \cite{Castro:2009ac}
in studying decay of D0-D6 bound states. 
For small black holes  the central charge does vanish, but  as we will see later, this occurs 
at rational and real values of the axion dilaton moduli not in the interior of the moduli space. 
If the central charged does vanish in the interior of the moduli space, then the corresponding 
state is not BPS. 

\begin{figure}
 \begin{center}
\epsfig{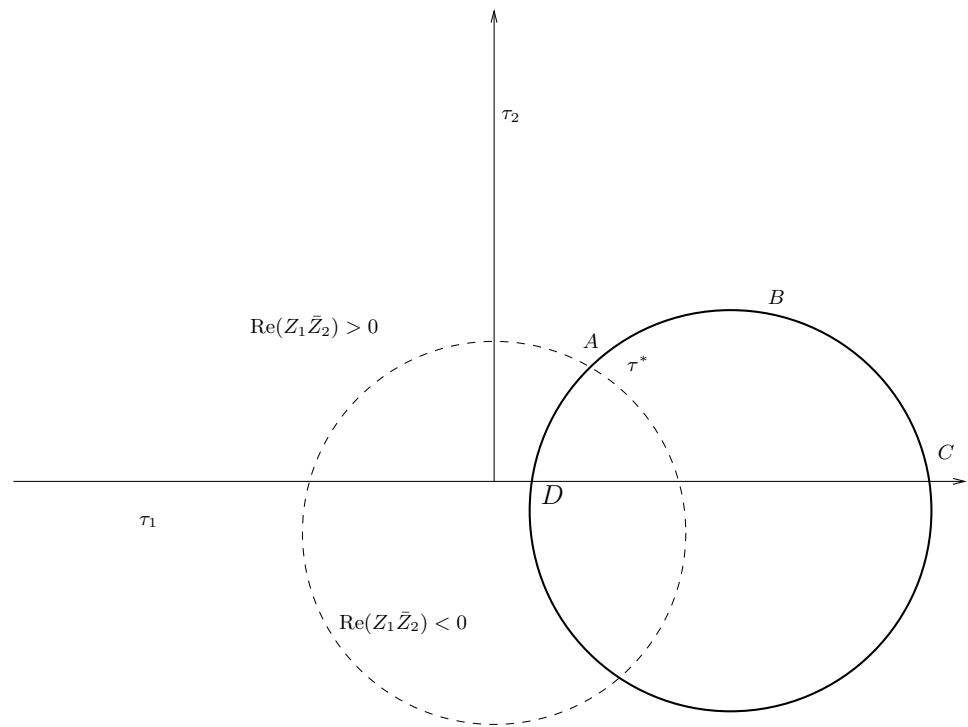}
\caption{Contradition obtained if the central charge $Z_1$ vanishes
in the interior of the moduli space: The inequalities  indicate the central charges $Z_1$ and $Z_2$
are anti-aligned along $DA$. But since the two black holes are at infinite distance 
along $DA$ the masses just add, therefore the central charges must be aligned.}
\label{fig-contra}
\end{center}
\end{figure}

In this section   we provide an alternate argument that the central charges do not vanish in the interior 
of the moduli space based on the continuity properties of the walls  of marginal stability. 
This will also enable use to derive the fact that the sign of 
${\rm{ Im}} (Q\cdot w P\cdot \bar w ) $
is maintained through out the moduli space. 
Let us suppose the central charge say $Z_1 = Z(Q_1, P_1)$ vanishes in the interior
of the moduli space. Consider any decay in which $Z_1$ is one of the decay products, we then 
have
\begin{equation}
 Z(Q, P)  = Z(Q_1, P_1)  + Z(Q_2, P_2).
\end{equation}
The wall of marginal stability in the upper half $\tau$ plane  is determined by the 
circle (\ref{imcond}) and the inequality (\ref{recond}). As discussed earlier, the two circles
intersect at
(\ref{intersect}), that is either at $Z_1=0$ or $Z_2=0$. Now that since the point $Z_1=0$  
lies in the interior of the moduli space, we will  have a situation which is schematically shown
in  figure \ref{fig-contra}. 
In this figure the circle with the bold line corresponds to the equation 
${\rm{Im}}(Z_1\bar Z_2) =0$, while the circle with the 
the dashed lines correspond to the equation ${\rm{Re}}(Z_1\bar Z_2) =0$.  They intersect at 
point $A$. For definiteness let the exterior of the circle with dashed lines be the domain 
in which ${\rm{Re}}(Z_1\bar Z_2) >0$.
Then the arc  $ABC$ is the segment along which both the 
equation (\ref{imcond}) and the inequality (\ref{recond}) are satisfied. 
Along the arc $DA$ the equation (\ref{imcond}) is satisfied while the 
inequality is not.  Therefore along the arc $ABC$ the charges $Z_1$ and $Z_2$ are aligned,
while along the arc $DA$, the charges are anti-aligned.
 Now the equation for the stability of the two centered black hole is 
given by (\ref{distance}). From this we see that two centered black hole solution will be 
stable at one side of the circle with bold lines. Let us assume that for definiteness it is 
stable in the interior of the circle with bold lines, that is (\ref{imcond}). 
This results in the following contradition: Consider a two centered black hole solution
along the arc $DA$, just in the interior on the bold circle.  The distance between 
the centres of the black holes from  (\ref{distance}) is infinite. Therefore they do not interact with 
each other.  Since they don't interact the total mass of the system is just sum of the 
masses of the two black holes \footnote{The author thanks
Ashoke Sen for  pointing this important fact.}. Therefore the mass of the corresponding 
single centred black hole  from which the two centred 
black hole has decayed  is 
given by 
\begin{equation}
\label{contra-1}
 |Z(Q, P)| = |Z(Q_1, P_1)| + |Z(Q_2, P_2)|.
\end{equation}
But from the fact along the arc $DA$ we have the condition ${\rm{Re}} (Z_1\bar Z_2) <0$,
the central charges are anti-aligned along $DA$. Thefore we have the equation
\begin{equation}
\label{contra-2}
 |Z(Q, P)| = |  |Z(Q_1, P_1)|- | Z(Q_2, P_2)| |.
\end{equation}
From equations  (\ref{contra-1}) and (\ref{contra-2}) we have obtained a contradition. 
Thus we see that the central charges cannot vanish in the interior of moduli space.  A
similar contradiction can be obtained if the two centered black hole is stable in the exterior 
of the circle  with bold lines and also if ${\rm{Re}}(Z_1\bar Z_2) > 0$ is satisfied in the 
interior of the circle with dashed lines. 

As mentioned earlier, there are cases in which the 
central charge can vanish at the boundaries of moduli space.
These occur for small black holes. Consider the case in  which the electric and the 
magnetic charges are proportional $(Q, P)= (m M, n M)$, Then the central charge vanishes at
$\tau = \frac{m}{n}$ with $M\cdot w \neq 0$. 
This point is at the boundary of moduli space,  one can 
map it to infinity by a $SL(2, \ZZZ)$ transformation. 
It is easy to see that the above argument
does not apply to such states, since the point $A$ is on the real line for these states not
in the interior of the moduli space. 
In fact these states are dual to purely electric states with electric charge $M$. The condition
$M\cdot w\neq 0$ restricts the argument from being applied to 
gauge bosons which can become massless at special points in moduli space.  Gauge bosons
satisfy the condtion $Q^2 =-1, P=0$, which are excluded from our analysis.

\vspace{.5cm}
\noindent
{\bf {Sign of ${\rm {Im}}( Q\cdot w P\cdot \bar w ) $}}
\vspace{.5cm}

Now that we know the central charge of any BPS state 
 does not vanish in the interior of the moduli space we can show that the 
the condition
$$
{\rm{ Im}} [(Q\cdot w ) (P\cdot \bar w) ] <0
$$ 
holds through in the interior of the moduli space. 
Let us suppose that there are regions in the moduli space where the above condition 
is violated. Let $\tilde w$ be in such a region, then 
\begin{equation}
\label{violated}
{ \rm{ Im}} (Q\cdot \tilde  w ) (P\cdot \bar{ \tilde w})  >0
\end{equation}
Let us examine the central charge at the point
\begin{equation}
 \tilde \tau_2 = \frac{ {\rm{ Im}} (Q\cdot \tilde w ) (P\cdot \bar {\tilde w})  }{ |P\cdot \tilde w|^2}, \qquad
\tilde \tau_1 =  \frac{ {\rm{ Re}} (Q\cdot \tilde w ) (P\cdot \bar {\tilde w})  }{ |P\cdot \tilde w|^2}.
\end{equation}
This point certainly lies in the interior of the moduli space due to \eq{violated}. 
At this point the central charge
\begin{equation}
 Z (Q, P ; \tilde \tau )  = ( Q\cdot \tilde w -  \tilde \tau  P\cdot \tilde w ) =0.
\end{equation}
We  have arrived at a contradition. Therefore there are no regions in the 
interior of moduli space
for which the condition  ${\rm{ Im}} [(Q\cdot w ) (P\cdot \bar w)]  <0$ is violated.
Thus once the branch \eq{finalform1} for the BPS mass formula is chosen, it does not change over the 
entire moduli space.

\section{Properties of  walls of marginal stability}

To get an idea of the geometric structure of the domains bounded by the various 
lines of marginal stability we need to study the possible intersection 
points of these walls.
We first consider two possible decays of a given charge vector $(Q, P)$
corresponding to a BPS state \footnote{$(Q,P)$ belong to the 
Narain lattice of the respective ${\cal N} =2$ model. We also choose 
to work in some duality frame in which  $P\neq 0$}.  
We denote the decays by the following equations
\begin{equation}
Z(Q, P) = Z(Q_1, P_1) + Z(Q_2, P_2), \qquad  Z(Q, P) = Z(Q_1', P_1') + Z(Q_2', P_2').
\end{equation}
For ease of notation we define
\begin{equation}
 Z_1 = Z(Q_1, P_1), \quad Z_2 = Z(Q_2, P_2), \quad   Z_1' = Z(Q_1', P_1'), \quad Z_2' = Z(Q_2', P_2').
\end{equation}
The equations that determine the two  walls of marginal stability are given by
\begin{eqnarray}
\label{twodecays}
 {\rm{Im }}( Z_1\bar Z) =0,  \qquad  {\rm{Re}} (Z_1\bar Z) >0, \\ \nonumber
{\rm{Im} } (Z_1' \bar Z) =0, \qquad { \rm{ Re}} (Z_1'\bar Z) >0.
\end{eqnarray}
Here we have used the equivalent form given in \eq{equivmargin} to write the  equations
for the walls of marginal stability.
From our earlier discussion, 
the walls of marginal stability of each of the two decays are determined
by the equality together with the restriction obtained from the
corresponding inequality.
We will examine these walls as sections in the $\tau$ plane
for a given moduli $w$. Therefore the physically relevant part of the
wall is that part which lies in the domain $\tau_2\geq 0$. 

We organize this section as follows:
In section 3.1 we show  that all walls of marginal decay of a given charge
vector $(Q, P)$ meet at a point in the lower half $\tau$ plane
( $\tau_2 <0$).  We then discuss the structure of 
the walls due to the decay of a given charge vector to two small black holes.
In section 3.3 we show that a wall can be characterized by  two 
real numbers $r_+$ and $r_-$ at which the circle intersects the real axis in the 
$\tau$ plane.   In section 3.4  we study  walls formed  by a generic black hole decay and 
state the conditions on the moduli and the charges which ensure 
a given decay physical. 
In section 3.5  we use the fact that all walls meet at a point in the lower half of the 
$\tau$ plane to  to determine the intersection 
point of a wall corresponding to a decay to small black holes and 
any generic decay. 
In section 3.6 we show that the only walls which don't intersect each other 
in the interior of the $\tau$ plane are the ones corresponding to the 
small black hole decay or the ones whose walls coincide with the 
walls of small black hole decays.
In section 3.7 we isolate the conditions on the moduli and the charges of 
a generic decay so that the corresponding wall always lies in a region
bounded by the walls corresponding to small black hole decays.

\subsection{All walls meet at the same point in the lower half $\tau$ plane}

We know that the equations  ${\rm{Im }}( Z_1\bar Z) =0$
and ${\rm{Im }}( Z_1'\bar Z) =0$  are generically circles. 
Therefore they intersect each other 
at the most twice. 
We now observe that a point common to these circles is the point $\tau^*$ where
$Z$ vanishes.  This point is given by 
\begin{equation}
\label{common-pt}
 \tau^*_1 =  \frac { {\rm{Re}}( Q\cdot w P\cdot \bar w) }{|P\cdot w|^2}, 
\qquad
\tau^*_2 =  \frac{ {\rm{Im}}( Q\cdot w P\cdot \bar w) } {|P\cdot w|^2}.
\end{equation}
 Since  ${\rm{Im}}( Q\cdot w P\cdot \bar w)<0$
for all $w$, this point lies in the lower half $\tau$ plane. 
  Thus we conclude 
that all walls corresponding to decays of a given 
 charge vector $(Q, P)$ meet at the point (\ref{common-pt})  in the lower half $\tau$ plane.
Though this point is not physically relevant 
 we will see that it provides us  useful information about the 
geometric structure of the domains formed by the walls.
In fact from the knowledge of this point it is easy to determine the possible second 
intersection point of the walls for a generic pair of decays. 
It is possible for the second point to be in a valid region of the moduli space and 
also satisfy the inequality  in \eq{twodecays}.
As we will see this second point determines the structure of the domains formed by the various
walls. 

Let us now restrict our attention  to the situation in which 
 one of the decays is to two small black holes in \eq{twodecays} 
and the other decay is  more generic, including decays to large black holes.
We start with examining  the wall corresponding to the decay to 
small black holes.

\subsection{Walls for decay into two small black holes}

By small black holes we mean those BPS states in these ${\cal N}=2$ theories
whose electric and magnetic charges are proportional. 
The charge vector $(Q, P)$ is given by $(m M, n M)$ where 
$M$ is a given vector in the Lorentzian lattice and $m, n$  are integers
\footnote{There can be other classes of small black holes, for instance
in the STU model, one can look at states whose electric and magnetic
charges are proportional in the $T$ frame rather than the 
usual $S$ frame. Our discussion will apply to these decays if 
one examines the wall in the $y^+$ plane for a given 
$\tau, y^-$ moduli. }.
We first study the lines of marginal stability for decay to  two small black holes.
We can then parametrize the  decay as  in \cite{Sen:2007vb}
\begin{equation}
\label{sen-parameters}
 \left(\begin{array}{c}
        Q\\P
       \end{array}
\right) =
\left(\begin{array}{c}
       ad Q - ab P\\ cdQ -cbP
      \end{array}
\right)
+ \left( \begin{array}{c}
-bc Q + ab P \\ -cd Q + ad P
\end{array}
\right),
\end{equation}
where $ad -bc =1$ and $\{a, b, c, d \} \in \ZZZ$. For simplicity we have assumed that the 
S-duality symmetry of the theory is $SL(2, \ZZZ)$. In the freely acting orbifold construction
of these theories  given in \cite{Ferrara:1995yx,Sen:1995ff, Gregori:1999ns,Gregori:1998fz}
the S-duality group is usually a subgroup of $SL(2, \ZZZ)$. In that case quantization of the charges leads to 
the condition:
$\{a, b,  d \} \in \ZZZ$ while $c \in N\ZZZ$ for  the 
S-duality group  $\Gamma_1(N)$ \cite{Sen:2007vb}.  One can easily generalize the conclusions 
found in this paper for these cases. 
As shown in \cite{Sen:2007vb}, the above parametrization is  a unique parametrization of the 
decay into two small black holes up to the following transformations
\begin{eqnarray}
\left( 
 \begin{array}{cc}
a & b \\ c & d 
\end{array} \right)
  \rightarrow 
\left( 
 \begin{array}{cc}
a & b \\ c & d 
\end{array} \right) \left( \begin{array}{cc}
\lambda & 0 \\
0 & \lambda^{-1} 
\end{array} \right) , \qquad
\left( 
 \begin{array}{cc}
a & b \\ c & d 
\end{array} \right)
\rightarrow 
\left( 
 \begin{array}{cc}
a & b \\ c & d 
\end{array} \right) \left( \begin{array}{cc}
0  & 1 \\ 
-1  & 0 
\end{array} \right).
\end{eqnarray}
Substituting the parametrization given in \eq{sen-parameters}
 into the conditions for marginal stability \eq{imcond} and \eq{recond}, 
 we obtain the following equation and inequality.  
\begin{eqnarray}
\label{sbh-decay}
& & {\rm{Im}}( Q\cdot w P\cdot \bar w) \left( 
cd |\tau|^2  - ( bc+ ad) \tau_1  + ab \right)  \\ \nonumber
& & - ( cd |Q\cdot w |^2 + ab |P\cdot w|^2 - (ad + bc) {\rm {Re}}( Q\cdot w P\cdot\bar w)  ) \tau_2 =0, 
\\ \nonumber
& & \left( -cd|Q\cdot w|^2 - ab |P\cdot w|^2 + (bc + ad) {\rm{Re}}( Q\cdot w P\cdot\bar w) \right)
\left( cd |\tau|^2 - ( ad + bc) \tau_1  + ab \right)  \\ \nonumber
& & - {\rm{Im}}( Q\cdot w P\cdot \bar w) \tau_2 >0.
\end{eqnarray}
It is convenient to perform a duality transformation to convert the conditions of marginal stability 
to straight lines. 
We consider the following S-duality transformations
\begin{equation}
\label{def-stau}
 \tau = \frac{a\tau' + b}{c\tau' + d}, \qquad \left(
\begin{array}{c}
 Q\\P
\end{array}
\right)
= \left(
\begin{array}{c}
 a\tilde Q + b \tilde P\\ c\tilde Q + d \tilde P
\end{array} \right)
\end{equation}
Then in terms of these new variables the equations reduce to 
\begin{eqnarray}
\label{sbh-line}
 -{\rm{Im}} (\tilde  Q\cdot w \tilde P\cdot \bar w) \tau_1' + 
{\rm{Re}} ( \tilde Q\cdot w \tilde P\cdot \bar w)\tau_2' =0\\ \nonumber
- {\rm{Re}} ( \tilde Q\cdot w \tilde P\cdot \bar w)\tau_1' - 
{\rm{Im}} ( \tilde Q\cdot w \tilde P\cdot \bar w) \tau_2' >0.
\end{eqnarray}
Now substituting for $\tau_1'$ from the first equation into the second inequality one 
obtains
\begin{eqnarray}
\label{inequality-sbh}
- \frac{|\tilde Q\cdot w \tilde P\cdot w|^2}{{\rm{Im}}( \tilde Q\cdot w \tilde P\cdot\bar w)} \tau_2' >0.
\end{eqnarray}
Since we have to examine the wall only in the 
physical $\tau$ plane, we have $\tau_2'\geq 0$. 
This implies that  the phases align on the line of marginal stability
only if ${\rm{Im}}( \tilde Q\cdot w \tilde P\cdot\bar w)<0$. 
It is easily seen that this condition is indeed true, since
\begin{eqnarray}
\label{basic-ineq}
 {\rm{Im}} [\tilde Q \cdot w \tilde P\cdot\bar w] &=& 
{\rm{Im}} [ ( dQ -bP)\cdot w ( -cQ + aP)\cdot \bar w ] , \\ \nonumber
&=& {\rm{Im}} (Q \cdot w P\cdot\bar w) <0.
\end{eqnarray}
Where we have substituted for $(\tilde Q, \tilde P) $
in terms of $(Q, P)$ from \eq{def-stau} and used the condition $ad -bc =1$. 
As we have shown that the the condition ${\rm{Im}} (Q \cdot w P\cdot\bar w) <0$ is maintained 
throughout the interior of the moduli space, 
the inequality (\ref{inequality-sbh}) is always satisfied in the 
upper half $\tau$ plane.   This  implies that the 
part of the line given in (\ref{sbh-line}) in the upper half plane 
is the wall of marginal stability for the decay of the charge vector
$(Q, P)$ to small black hole given by equation (\ref{sen-parameters}) 
{\emph {Therefore given  the charge vector $(Q, P)$ and the moduli $w$ there  always  exists 
a wall of marginal stability for the charge vector to decay into small black holes. }}

The analysis of the structure of the domains formed by these walls was done in 
\cite{Sen:2007vb} for ${\cal N}=4$ theories. The same analysis goes through for these class of 
${\cal N}=2$ models. 
A schematic diagram of the domains is given in figure 1. The small 
black hole decays  consists of the bold curves. There are lines 
 passing through the integer points on the real axis and there are 
circles joining each of these points. Then there  are other circles  which always
lie below the circles joining the integer points.
None of the circles corresponding to small black hole decay intersect each other in the interior of the 
upper half plane.  The circles can intersect in the physical part of the $\tau$ plane only on the 
real axis. 
From \eq{sbh-decay} we see that the wall of marginal stability intersects the 
real axis in the $\tau$ plane at 
\begin{equation}
\label{sbh-intersect}
r_+ = \frac{a}{c}, 
\quad \hbox{and}, \quad r_- = \frac{b}{d}, \quad\hbox{with}\quad ad-bc =1,
\end{equation}
 and it is only these 
points that can be possible meeting points of the walls corresponding to two different 
small black hole decays.

\subsection{Characterization of a generic decay}

We have seen in the previous section that small black holes decays are characterized by 
the integers $a, b, c, d$ with $ad-bc =1$. In this section we introduce 
a simple method of characterization of the wall for a general decay.. 
This characterization  enables one to easily determine if 
two walls intersect each other in the interior of the upper half plane.
As we will see, this enables one to easily classify walls. 
Consider a generic decay given by
\begin{equation}
\label{gen-decay}
 \left(\begin{array}{c}
        Q\\ P 
       \end{array}
\right) =  
 \left(\begin{array}{c}
        Q_1'\\ P_1' 
       \end{array}
\right)
+
 \left(\begin{array}{c}
        Q_2'\\ P_2' 
       \end{array}
\right)
\end{equation}
From \eq{imcond} and \eq{recond} it is easy to see that the
wall of marginal stability of this decay is determined by the following
\begin{eqnarray}
\label{circlelb1}
 & & \tau\bar\tau  {\rm Im} [ P_1'\cdot w P \cdot \bar w] - 
\tau_1  {\rm Im} [ (P_1' \cdot w  Q \cdot \bar w)  +( Q_1'\cdot w P \cdot \bar w)]   \\ \nonumber
& & -\tau_2 {\rm Re} [(P_1' \cdot w  Q \cdot \bar w)  -( Q_1'\cdot w P \cdot \bar w)] 
+ {\rm Im} [ Q_1'\cdot w  Q\cdot \bar w ] =0    \\
\label{circlelb2}
& & \tau\bar\tau {\rm Re} [ P_1'\cdot w P_2' \cdot \bar w]
- \tau_1 {\rm Re} [ (P_1' \cdot w  Q_2' \cdot \bar w)  +( Q_1'\cdot w P_2' \cdot \bar w)] \\ \nonumber
& & + \tau_2 {\rm Im} [(P_1' \cdot w  Q_2' \cdot \bar w)  -( Q_1'\cdot w P_2' \cdot \bar w)]
+ {\rm Re} [ Q_1'\cdot w  Q_2'\cdot \bar w ]  >0
\end{eqnarray}
We now wish to state the conditions under which the above wall of marginal stability 
is physical. That is, the conditions that 
ensure that  the  part of the circle in \eq{circlelb1} lies in the upper half $\tau$-plane 
and this part of the circle satisfies the inequality in \eq{circlelb2}.
For ease of notation let us again define the coefficients
\begin{eqnarray}
\label{ABC-coef}
A = {\rm Im} [ P_1'\cdot w P \cdot \bar w], 
& \quad&  A'  = {\rm Re} [ P_1'\cdot w P_2' \cdot \bar w], \\ \nonumber
B =  {\rm Im} [ P_1' \cdot w Q \cdot \bar w  + Q_1'\cdot w P\cdot \bar w] ,
& \quad&  B' = {\rm Re } [ P_1' \cdot w  Q_2' \cdot \bar w  + Q_1'\cdot w P_2'\cdot \bar w], \\ \nonumber
C = {\rm Re} [P_1' \cdot w  Q \cdot \bar w  - Q_1'\cdot w P \cdot \bar w],
& \quad&  C' = - {\rm Im} [P_1' \cdot w  Q_2' \cdot \bar w  - Q_1'\cdot w P_2' \cdot \bar w ], \\ \nonumber
D = {\rm Im} [ Q_1'\cdot w  Q\cdot \bar w],
& \quad&  D' = {\rm Re} [ Q_1'\cdot w  Q_2'\cdot \bar w ].
\end{eqnarray}
We already know that the circle \eq{circlelb1} passes through the point $\tau^*$  given in \eq{common-pt} 
which lies in the lower half $\tau$ plane. Therefore, the necessary  condition one needs to impose so that
part of the circle in \eq{circlelb1} lies in the upper half plane is that it should intersect the 
real axis. This is given by the following 
\begin{eqnarray}
\label{no-intersect}
& &  B^2 - 4 AD >0, \quad {\rm{or}} \\ \nonumber
& & ( {\rm Im} [ P_1' \cdot w Q \cdot \bar w  + Q_1'\cdot w P\cdot \bar w] )^2 
- 4 {\rm Im} [ P_1'\cdot w P \cdot \bar w] {\rm Im} [ Q_1'\cdot w  Q\cdot \bar w] >0.
\end{eqnarray}

We also must make sure that if \eq{circlelb1} is to be a physical 
line of marginal stability, the points on the curve in \eq{circlelb1} must also satisfy the 
inequality \eq{circlelb2}. We will discuss the various conditions which ensure this in
the next section.  We now introduce a method to characterize the 
circles in \eq{circlelb1} provided it satisfies the condition \eq{no-intersect}.  
Instead of choosing $A, B, C, D$  to specify this circle we 
characterize the circle as follows: 
We already know that this circle passes through the common point 
\eq{common-pt}. This point is completely specified by the initial 
charge vector $(Q, P)$ and the moduli $w$. 
We need two more points to characterize this circle. 
Due to the condition \eq{no-intersect}, it is clear that it intersects the real axis. 
The points of intersection are given by
\begin{equation}
r_{\pm } = \frac{ B \pm \sqrt{ B^2  - 4 AD }}{A}.
\end{equation}
Without loss of generality we will assume $r_- < r_+$, the case when 
$r_- =r_+$ just results in the circle being tangent to the real line from
below. 
Since the  point \eq{common-pt} is common to all possible decays,
specifying   the two points of intersection 
on the real axis $r_+$ and $r_-4$ completely determines the circle. 
Note that when $A\rightarrow 0$, $r_+\rightarrow \infty$, then the circle reduces to 
a line and $r_-$ then refers to the point of intersection of the line with the real axis.
Though one has uniquely specified the circle using these numbers, the decay
which corresponds to a circle specified by $r_-$ and $r_+$ is not unique. 
 A decay  with 
the parameters $\lambda A, \lambda B, \lambda C,  \lambda D$, 
will also have the same values of $r_+$ and $r_-$ and pass through
\eq{common-pt}. The advantage of this characterization of the circle is
that given a pair of circles $(r_-, r_+)$ and $(r_-', r_+')$ they intersect
in the interior of the upper half plane if and only if either of the following 
conditions are satisfied
\begin{equation}
\label{int-intersect}
r_- < r_-' < r_+ < r_+', \quad\hbox{or} \quad r_-'< r_- <r_+' < r_+.
\end{equation}

\subsection{Existence  conditions for walls of  generic decays}

In this section we will find the conditions necessary so that 
 the part of the circle \eq{circlelb1} which emerges in the 
upper half plane also satisfies the inequality \eq{circlelb2}.
Let us examine the situation when $A= {\rm{ Im} }(P_1'\cdot w P\cdot \bar w) \neq 0$. 
Then we can use the equation in \eq{circlelb1} to write the inequality in 
\eq{circlelb2} as 
\begin{equation}
\label{line-ineq}
\frac{ A'B - A B'}{A} \tau_1 + \frac{( A'C - AC')}{A} \tau_2 + \frac{( AD' - A'D)}{A}  >0.
\end{equation}
We also know from the discussion above \eq{intersect} that the line which determines the inequality 
in \eq{line-ineq} intersects the circle \eq{circlelb1} only in the lower half $\tau$ plane. 
This is because the intersection of the line in \eq{line-ineq} and the circle \eq{circlelb1} 
occurs at points where the central charges $Z_1'$ or $Z_2'$ vanishes and these 
are only in the lower half $\tau$ plane. 
Therefore, there are no other points at which the line \eq{line-ineq} intersects the 
circle \eq{circlelb1}.  This implies that to ensure that the part of the circle 
which emerges in the upper half plane satisfies the inequality it is sufficient to  
demand that  any point  on the circle \eq{circlelb1} satisfies
the inequality \eq{line-ineq}.
We know that $\tau = (r_+, 0)$ and $\tau = (r_- , 0)$ are points on the circle \eq{circlelb1}. 
Thus the average of these points also must satisfy the inequality \eq{line-ineq}. 
This gives the condition
\begin{eqnarray}
\label{phys-wall1}
( r_+ + r_-) \frac{A'B - AB'}{A}  + 2\frac{AD' - A'D}{A} >0, \\ \nonumber
B(A'B -AB')   + 2( AD' - A'D)A >0,
\end{eqnarray}
where we have used $r_+ + r_- = B/A$. 
Note that this true only if $A'B -AB' \neq 0$. 
Rewriting the above condition in terms of the values of the coefficients
and after some simplifications we obtain
\begin{eqnarray}
\label{phys-wall}
& & \left[ |P_1'\cdot w |^2 {\rm Im}( P_2'\cdot w Q_2'\cdot \bar w) +
 | P_2\cdot w|^2 {\rm Im}(  P_1'\cdot w Q_2 '\cdot \bar w)   \right]  \\ \nonumber
& & \times {\rm Re} ( P_1' \cdot  w 
Q_2 \cdot \bar w + Q_1' \cdot w P_2' \cdot \bar w ) 
 -2 {\rm } (P_1' \cdot w P_2'\cdot  \bar w ) {\rm Im} ( P_1'\bar w  P_2 \cdot w Q_1'\cdot w Q_2\cdot \bar w) 
>0.
\end{eqnarray}
For completeness let us discuss the case in which $A'B -AB'=0$, evaluating this explicitly we obtain
\begin{equation}
 A'B -AB' = |P_1'\cdot w |^2 {\rm Im}( P_2'\cdot w Q_2'\cdot \bar w) +
 | P_2\cdot w|^2 {\rm Im}(  P_1'\cdot w Q_2 '\cdot \bar w),
\end{equation}
Since both the terms are of the same sign, 
this can vanish only if ${\rm Im}( P_2'w Q_2 \bar w ) = {\rm Im} ( P_1'\bar w Q_1'\bar w) =0$.
This occurs when both the decay products are small black holes. 
Now $$AD' -A'D = - {\rm Im} ( P_1'\bar w  P_2 \cdot w Q_1'\cdot w Q_2\cdot \bar w) $$
which also vanishes for small black hole decays. Thus the condition 
\eq{phys-wall1} for small black hole decay reduces to 
\begin{equation}
\frac{A'C - AC'}{A} >0 \quad \hbox{if}\quad A\neq 0..
 \end{equation}
We can verify this  condition for the existence of a physical wall of marginal 
for small black hole decay. 
Evaluating the condition \eq{no-intersect} for this case by substituting the values 
of the charges of the decay products from \eq{sbh-decay}  and evaluating the 
coefficients from either \eq{ABC-coef} or \eq{sbh-decay} we obtain 
\begin{equation}
 B^2 -4AD = [  {\rm{Im}} ( Q\cdot w P\cdot\bar w) ]^2 ( bc -ad )^2  >0,
\end{equation}
which is always satisfied. Furthermore we see that 
${\rm{ Im}} (P_1'\cdot w P\cdot \bar w) = cd {\rm{Im}}( Q\cdot w P\cdot\bar w) \neq 0$ if
$cd \neq 0$.   Now examining the condition in \eq{phys-wall} we see
\begin{equation}
 - \frac{ 
 \left\{ -cd |Q\cdot w|^2 - ab |P\cdot w|^2 +
  ( bc + ad ) {\rm {Re}}( Q\cdot w P\cdot\bar w) \right\}^2
 + [ {\rm Im} (  Q\cdot w P \cdot \bar w ) ]^2   }
 {{\rm{Im}} ( Q\cdot w P\cdot\bar w) }   >0 .
\end{equation}
It is now clear that the wall is physical when ${\rm{Im}} ( Q\cdot w P\cdot\bar w)<0$
which as we have seen is always true. 
Thus the decay to two small black holes is always allowed, which is the same conclusion reached earlier. 

Let us now  consider the case of   $A={ \rm{ Im}} (P_1'\cdot w P\cdot \bar w) =0$.
Note that this condition in general imposes an additional condition on the 
$w$ moduli. But then the wall of marginal stability we will obtain is a 
higher co-dimension  surface.  Therefore we must look for   situations 
when $A$ vanishes at generic $w$ moduli. This occurs if $P_1'=0$
or if $P_1' =\alpha P$ and $\alpha \neq 1$ \footnote{ $P\neq 0$ 
by assumption,
$P_1' =P$ is the same physical situation as $P_1'=0$, because in this case $P_2' =0$}. 
Let us first examine the situation when $P_1'=\alpha P$. 
 Now in this case, the circle in \eq{circlelb1} reduces to a line.
\begin{equation}
\label{line-case1}
-B\tau_1 -C\tau_2 + D =0,
\end{equation}
while the  coefficient $A'$ reduces to $\alpha(1-\alpha) |P\cdot w|^2 $.
We know the intersection of the circle which determines the inequality in 
\eq{circlelb2} and the line  in \eq{line-case1} lies in the lower half plane. 
If the points on the line in \eq{line-case1} satisfies the inequality \eq{circlelb2}, then 
it must be true that the inequality must hold when $\tau_2\rightarrow\infty$. 
Thus if the coefficient $A'>0 $  the inequality \eq{circlelb2} is satisfied and 
 the resulting decay is physical. But since $A' = \alpha(1-\alpha) |P\cdot w|^2 $ 
we must have $0<\alpha<1$ for $A'>0$.  
Now let us examine the situation when $P_1=0$.
For this   case we see that 
the coefficient $A'$ also vanishes. Thus both the circle in \eq{circlelb1} and the circle 
determining the inequality \eq{circlelb2} reduces to following lines.
\begin{equation}
 -B\tau_1 - C\tau_2 + D =0, \qquad
-B'\tau_1 - C'\tau_2 + D' >0.
\end{equation}
Let us suppose $ B = {\rm{Im}} ( Q_1'\cdot w P\cdot \bar w) \neq 0$.
The lines intersect at a point  in the lower half plane. This is the point at which one of the 
central charges vanish. Eliminating $\tau_1$ from the first equation one gets the condition
\begin{equation}
\label{red-eq}
 \frac{B'C -BC'}{B} \tau_2 - \frac{B'D + B D'}{B} >0.
\end{equation}
The imaginary part of $\tau$ at which  the lines meet is  given by
\begin{equation}
 \tau_2^* = \frac{B' D - BD'}{B'C - BC'} = \frac{{{\rm Im}} ( Q_2' \cdot w P\cdot \bar w) }{|P\cdot w|^2} <0.
\end{equation}
Note that this is the point at which the central charge $Z_2'$ vanishes and it lies in the lower half
of the  $\tau$ plane. 
Therefore we  obtain the condition that the  wall of marginal stability is physical if and only if  the coefficient
of $\tau_2$ in  \eq{red-eq} is positive
\begin{equation}
\label{physcase2}
 \frac{B'C -BC'}{B} =\frac{  |Q_1'\cdot w|^2 {\rm{Im}} (P\cdot w Q_2'\cdot\bar w)  }
{ {\rm{Im}}( Q_1'\cdot w  P\cdot \bar w) } >0.
\end{equation}
Finally we look at the case $P_1'=0$ which implies $A=0$ and 
$ B = {\rm{Im}} ( Q_1'\cdot w P\cdot \bar w) = 0$.
For generic values of $w$ moduli, this can occur only when $Q_1' = \alpha P$ for 
$\alpha \neq 0$ since for $\alpha =0$, there is no decay. 
Then the equation of the circle \eq{circlelb1} reduces to 
\begin{equation}
\alpha |P\cdot w|^2 \tau_2 - \alpha {\rm{Im}}( Q\cdot w P\cdot \bar w) =0.
\end{equation}
Since ${\rm{Im}}( Q\cdot w P\cdot \bar w)<0 $, this equation has no solution
in the physical upper half $\tau$ plane.

We now summarize all the conditions 
necessary for the occurrence of a physical wall of marginal stability in the following table. 
For all the cases the condition in \eq{no-intersect}
is necessary. Then the conditions for the existence of a physical 
wall of marginal stability splits into the following  cases: \\
\begin{center}
\vspace{.3cm}
\begin{tabular}{l l l }
\hline
 &  \\ 
 Case (i)  & 
$A\neq 0, A'B - AB' \neq 0 $ and  & \\
 &  $  (A'B - A B')B  + 2(AD' -A'D)A >0 $ & See \eq{phys-wall} \\
 & \\
Case (ii)  & $A=0, P_1' =0, B, \neq 0 $  and $\frac{B'C -BC'}{B}>0$ & See \eq{physcase2} \\
 & $A=0, P_1'= \alpha P$ and $0 <\alpha <1$ , $\Rightarrow A'>0$,  &Decay allowed  \\ 
& \\
Case (iii)  & Small black hole decay    $A'B - AB' =0 $  & Always allowed
 \\
& \\
\hline
\end{tabular}\\
\vspace{.3cm}
{\small {\bf Table 1. }}{\small  Conditions for the existence for a wall of marginal stability }
\end{center}
In what follows we will assume that the moduli $w$ for this general decay always satisfies
these conditions.

\subsection{Intersection of walls of  small black hole decays and a generic decay}

Since a generic decay is a circle seen in the $\tau$ plane, the walls of two decays intersect at the most twice. 
We have seen that all walls meet at a common point in the lower half plane.
Using this fact it is easy to find the other possible point of intersection. 
In this section we will find this second point of intersection, for a wall corresponding to 
a small black hole decay and a generic decay. The generic decay can include decays
to large black holes. 
Let us denote the second decay by that given in \eq{gen-decay}.
To find the intersection point of the circle \eq{circlelb1} and that corresponding to 
the small black hole decay in \eq{sbh-decay}.  It is convenient to  parametrize the
upper half $\tau$ plane  using the coordinate $\tau'$ given in 
\eq{def-stau}. In this coordinate, the equation \eq{circlelb1} becomes
\begin{eqnarray}
\label{trans-circle}
 & & |\tau'|^2 \left\{ a^2 {\rm{ Im}} (P_1\cdot w P\cdot \bar w) 
-ac {\rm Im}( P_1\cdot w Q\cdot \bar w + Q_1 \cdot w P\cdot \bar w)
+ c^2 {\rm Im} ( Q_1\cdot w Q\cdot \bar w) \right\} \nonumber \\ 
& & \tau_1' \left\{ 
2ab {\rm{ Im}} (P_1\cdot w P\cdot \bar w) -
(ad + bc) {\rm Im}( P_1\cdot w Q\cdot \bar w + Q_1 \cdot w P\cdot \bar w) \right.  \\ \nonumber
& & \left. + 2cd {\rm Im} ( Q_1\cdot w Q\cdot \bar w) \right\}  
 - \tau_2' {\rm Re} ( P_1\cdot w Q\cdot \bar w - Q_1 \cdot w P\cdot \bar w) \\ \nonumber 
& & 
+ \left\{
b^2 {\rm{ Im}} (P_1\cdot w P\cdot \bar w)  - bd 
{\rm Im}( P_1\cdot w Q\cdot \bar w + Q_1 \cdot w P\cdot \bar w)  
+ d^2 {\rm Im} ( Q_1\cdot w Q\cdot\bar w) \right\}
=0.
\end{eqnarray}
While the small black hole decay reduces to the line given in 
\eq{sbh-line}.
From the discussion in the earlier section 3.1  and from  
\eq{common-pt} we know that, there is always the point 
\begin{equation}
 \tau^{*\prime}  = \frac{Q'\cdot w}{P'\cdot w} = \frac{ d \tau^* - b}{-c\tau^* + a},
\end{equation}
at which the line  in \eq{sbh-line} and the circle in \eq{trans-circle} meet. 
Here we have just rewritten the point in \eq{common-pt} in the $\tau'$ co-ordinates. 
To find the other point we first substitute
\begin{equation}
\label{eqofline}
 \tau_1' = \frac{{\rm Re} ( \tilde Q\cdot w \tilde P \cdot \bar w) }
{ {\rm Im} (\tilde Q\cdot w \tilde P\cdot \bar w) } \tau_2',
\end{equation}
which arises from the \eq{sbh-line} into \eq{trans-circle} to obtain a quadratic 
equation for the points of intersection.
This equation is given by
\begin{eqnarray}
& &  \left(
\frac{ |\tilde Q\cdot w \tilde P\cdot \bar w|^2}{ ( {\rm{Im}}( \tilde Q\cdot w \tilde P\cdot \bar w) ) ^2 }
\right)
\left\{ a^2 {\rm{ Im}} (P_1'\cdot w P\cdot \bar w) 
-ac {\rm Im}( P_1'\cdot w Q\cdot \bar w + Q_1' \cdot w P\cdot \bar w)
 \right.\ \nonumber \\ 
& & \qquad \qquad \qquad\qquad\qquad
\left. +   c^2 {\rm Im} ( Q_1'\cdot w Q\cdot \bar w) \right\}  \tau_2^{\prime 2}
\\ \nonumber
& & + \left\{
\frac{{\rm{Re}} (\tilde Q\cdot w \tilde P\cdot \bar w)}{ { \rm{Im}} (\tilde Q\cdot w \tilde P\cdot \bar w )} 
\left[ 2ab {\rm{ Im}} (P_1'\cdot w P\cdot \bar w) -
(ad + bc) {\rm Im}( P_1'\cdot w Q\cdot \bar w + Q_1' \cdot w P\cdot \bar w) \right. \right.   \\ \nonumber
& & \qquad\qquad\qquad\qquad\qquad  \left. \left. + 2cd {\rm Im} ( Q_1'\cdot w Q\cdot \bar w) \right]
- {\rm Re} ( P_1'\cdot w Q\cdot \bar w - Q_1' \cdot w P\cdot \bar w) \right\}
\tau_2^{\prime} 
\\ \nonumber
& & +  \left\{ b^2 {\rm{ Im}} (P_1'\cdot w P\cdot \bar w)  - bd 
{\rm Im}( P_1'\cdot w Q\cdot \bar w + Q_1' \cdot w P\cdot \bar w)  
+ d^2 {\rm Im} ( Q_1'\cdot w Q\cdot \bar w) \right\}
= 0.
\end{eqnarray}
From the above equation, one can easily read out the product of the two roots,
which  is given by
\begin{eqnarray}
 \label{prod-roots}
& & \frac{ 
 b^2 {\rm{ Im}} (P_1'\cdot w P\cdot \bar w)  - bd 
{\rm Im}( P_1'\cdot w Q\cdot \bar w + Q_1' \cdot w P\cdot \bar w)  
+ d^2 {\rm Im} ( Q_1'\cdot w Q\cdot \bar w) 
}{
 a^2 {\rm{ Im}} (P_1'\cdot w P\cdot \bar w) 
-ac {\rm Im}( P_1'\cdot w Q\cdot \bar w + Q_1' \cdot w P\cdot \bar w)
+ c^2 {\rm Im} ( Q_1'\cdot w Q\cdot \bar w)  } \nonumber \\ 
& &  \qquad\qquad\qquad\qquad\times
\frac{ ( {\rm{Im}} ( \tilde Q\cdot w \tilde P\cdot \bar w) )^2}{ |\tilde Q\cdot w \tilde P\cdot \bar w|^2 }.
\end{eqnarray}
Therefore the second root of the quadratic equation is given by
\begin{eqnarray}
\label{2-intersect}
\tilde \tau_2^\prime &=& \frac{ 
 b^2 {\rm{ Im}} (P_1'\cdot w P\cdot \bar w)  - bd 
{\rm Im}( P_1'\cdot w Q\cdot \bar w + Q_1' \cdot w P\cdot \bar w)  
+ d^2 {\rm Im} ( Q_1'\cdot w Q\cdot \bar w) 
}{
 a^2 {\rm{ Im}} (P_1'\cdot w P\cdot \bar w) 
-ac {\rm Im}( P_1'\cdot w Q\cdot \bar w + Q_1' \cdot w P\cdot \bar w)
+ c^2 {\rm Im} ( Q_1'\cdot w Q\cdot \bar w)  } \nonumber \\
& &
\times
\frac{  {\rm{Im}} ( \tilde Q\cdot w \tilde P\cdot \bar w)   }{ |\tilde Q\cdot w |^2 }.
\end{eqnarray}
To obtain the second root we have divided the product of the roots by the 
imaginary part of 
$\tau_2^{*\prime}$.
Then using the equation of the line \eq{eqofline} we can find the real part of the
the second point of intersection. This is given by
\begin{eqnarray}
\tilde{\tau}_1 ^\prime &=&
\frac{ 
 b^2 {\rm{ Im}} (P_1'\cdot w P\cdot \bar w)  - bd 
{\rm Im}( P_1'\cdot w Q\cdot \bar w + Q_1' \cdot w P\cdot \bar w)  
+ d^2 {\rm Im} ( Q_1'\cdot w Q\cdot \bar w) 
}{
 a^2 {\rm{ Im}} (P_1'\cdot w P\cdot \bar w) 
-ac {\rm Im}( P_1'\cdot w Q\cdot \bar w + Q_1' \cdot w P\cdot \bar w)
+ c^2 {\rm Im} ( Q_1'\cdot w Q\cdot \bar w)  } \nonumber \\ 
& &
\times
\frac{  {\rm{Re}} ( \tilde Q\cdot w \tilde P\cdot \bar w)   }{ |\tilde Q\cdot w |^2 }.
\end{eqnarray}
In general the second  point of intersection 
between a small black hole decay and a generic 
decay can be in the interior of the  upper half plane if $\tilde \tau_2'> 0$.

\subsection{Walls which never intersect in the interior of the moduli space}

In the following we will find the
necessary and sufficient
 conditions such that this second point $\tilde \tau'$  never lies 
 in the interior of the upper half plane  given
 any small black hole decay specified by $a, b, c, d, \in \ZZZ$ with $ad -bc =1$. 
From \eq{2-intersect} and the fact that
  ${\rm{Im}}(  Q'\cdot w P'\cdot \bar w) <0$ for all $w$, 
 we see that  the only way 
  $\tilde\tau_2'\leq 0$,  is to demand
\begin{eqnarray}
\label{def-ratio}
 R&=& \frac{ 
\left( b^2 {\rm{ Im}} (P_1'\cdot w P\cdot \bar w)  - bd 
{\rm Im}( P_1'\cdot w Q\cdot \bar w + Q_1' \cdot w P\cdot \bar w)  
+ d^2 {\rm Im} ( Q_1'\cdot w Q\cdot \bar w) \right)
}{
\left( a^2 {\rm{ Im}} (P_1'\cdot w P\cdot \bar w) 
-ac {\rm Im}( P_1'\cdot w Q\cdot \bar w + Q_1'\cdot w P\cdot \bar w)
+ c^2 {\rm Im} ( Q_1'\cdot w Q\cdot \bar w) \right) }, \nonumber \\ &\geq& 0,
\end{eqnarray}
for all $a, b, c, d$ which satisfies $ad -bc=1$. 
We will now  show that there are two possible ways to ensure this.

Observe that numerator and the denominator that occurs in the ratio $R$  can be written as
\begin{equation}
 R = \frac{ v_2^T N v_2}{ v_1^T N v_1},
\end{equation}
where 
\begin{eqnarray}
 v_2 = \left(\begin{array}{c}
                b \\d 
               \end{array}
\right), \qquad\qquad
v_1 = \left( 
\begin{array}{c}
 a  \\ c
\end{array}
\right) \qquad 
\label{def-N}
N &=&  \left(\begin{array}{cc}
              A &  -\frac{B}{2}  \\
- \frac{B}{2}  &  D
             \end{array}
\right).
\end{eqnarray}
Now it is clear that if the product of eigen values of the matrix $N$ turns of to be positive then the ratio $R>0$ for all $a, b, c, d$. 
For this to be true we must have
the ${\rm {Det}}(N) >0$.  This results in 
\begin{equation}
\label{det-cond}
 \left[ {\rm {Im} }(P_1' w P\cdot \bar w) {\rm{Im}} ( Q_1'\cdot w Q\cdot \bar w) \right] ^2 
-\frac{1}{4} [ {\rm{Im}}( P_1'\cdot w Q\cdot\bar w + Q_1'\cdot w P\cdot \bar w) ]^2 >0.
\end{equation}
From our discussion  in the section 3.3
and from \eq{no-intersect} we see that the above condition 
ensures that the wall of marginal stability for the generic decay in \eq{circlelb1} never
intersects the real  axis. Further more since it always passes through the 
point \eq{common-pt} in the lower half $\tau$-plane, it lies entirely in the lower
half plane. 
We exclude this situation as the equation \eq{circlelb1} does not represent
a physical wall of marginal stability. 

Now if ${\rm{Det}}(N)< 0$,
one can factorize the 
quadratic form
\begin{equation}
 A x^2  -B  xy + D  y^2,
\end{equation}
where $A, B, D$ are defined in \eq{ABC-coef}.
We first examine the situation when  
$A ={\rm{Im}}( P_1'\cdot w P\cdot \bar w)  \neq 0$, we then can 
write the above quadratic form as 
\begin{equation}
 A ( x - r_{+} y) ( x - r_{-} y), \quad
\hbox{where}\quad
 r_{\pm } = \frac{ B \pm \sqrt{ B^2  - 4 AD }}{A}.
\end{equation}
Note that these roots are real as we have ${\rm{Det}}(N)< 0$ and these
are the points the the circle \eq{circlelb1} intersects the real axis. 
 The ratio $R$ for this situation can be written as
\begin{equation}
 R = \frac{ (b - r_+ d) ( b - r_- d) }{ ( a- r_+ c) ( a- r_-c) },
\end{equation}
We need this ratio $R\geq 0$ for all $a, b, c, d$ with $ad -bc =1$.
We first  show that this can be maintained for all $a, b, c, d$ such 
that $ad -bc =1$ if $r_+$ and $r_-$ are rational with the condition
\begin{equation}
\label{class-1}
 r_+ = \frac{p}{q}, \quad r_-=\frac{p '}{q'}, \;\hbox{and}\quad pq' - p'q = 1.
\end{equation}
 Here we have chosen $r_+>r_-$ for definiteness. One can interchange
the assignment if $r_- >r_+$. 
Consider the product of the matrices 
\begin{equation}
 \left(
\begin{array}{cc}
  q'   & - p' \\ - q & p 
\end{array}
\right) 
\left(\begin{array}{cc}
       a & b \\ c & d
      \end{array}
\right)  =
\left[\begin{array}{cc}
       ( a q' - p' c) & ( b q' - p' d) \\
( - a q + p c) & ( -bq + pd )
      \end{array}
\right].
\end{equation}
Note that since $pq'-p'q =1$, the first matrix in the above equation is a $SL(2,\ZZZ)$ matrix.  Since the product of two $SL(2, \ZZZ)$ matrices
 is also a $SL(2, \ZZZ)$ matrix we see that
\begin{equation}
 ( a q' - p' c) ( -bq + pd ) - ( b q' - p' d)( - a q + p c) =1.
\end{equation}
 Each of the terms in the above equation is an integer, we have the sign of 
$ ( a q' - p' c) ( -bq + pd )$ and $( b q' - p' d)( - a q + p c)$ is the same or either 
one of the terms is zero. This implies that the ratio
\begin{eqnarray}
 R &=&  \frac{( -bq + pd )( b q' - p' d) }{ ( - a q + p c)( a q' - p' c)},  \\ \nonumber
&=& \frac{ ( b - \frac{p}{q} d ) ( b - \frac{p'}{q'} d) }{
( a - \frac{p}{q} c ) ( a - \frac{p'}{q'} c) },  \\ \nonumber
&=&  \frac{ (b - r_+ d) ( b - r_- d) }{ ( a- r_+ c) ( a- r_-c) },  \\ \nonumber
&\geq&0,
\end{eqnarray}
for all $\{a, b, c,d \} \in \ZZZ$ with $ad -bc \in \ZZZ$. 
From the previous section 3.4, we see that if these decays have to be physical then
the coefficients  of \eq{circlelb1} and \eq{circlelb2} also must satisfy  Case (i) of Table 1. 
Note that this basically imposes and inequality on the coefficients $A'  B', D$. 
Since these are conditions on independent
coefficients $A',B', D'$, we can always work in the domain of $w$ such that  
the inequality in \eq{phys-wall} is satisfied. 
Note that  $r_+$ and $r_-$ are the points at which 
the circle \eq{circlelb1} intersects the real axis in the original $\tau$ plane
and they agree with the points that   the wall corresponding to 
some small black hole decay intersects
 the real axis given in \eq{sbh-intersect}. Therefore 
 according to our discussion in section 3.3  the decays which satisfy 
\eq{class-1} must coincide with the walls corresponding to some small 
black hole decay. It is important to mention that though the walls 
of such decays are the same as that of the small black hole decay, the 
actual decay can be different, since the values $r_+$ and $r_-$ do not 
characterize the decay uniquely but only the corresponding wall.

The case when ${\rm Det}(N) =0$ is not of sufficient interest because, in such a situation
$r_+ = r_ -  $ and the wall of marginal stability for the generic decay \eq{circlelb1} 
intersects the real line only once from below. It does not emerge in the upper half  $\tau$ plane and 
therefore not a physical wall. The only remnant of this wall in the physical $\tau$ plane is 
a point on the real axis.   

We finally  examine the situation when $A ={\rm{Im}}( P_1'\cdot w P\cdot \bar w)  =0$.
From table 1 we see that this occurs for $P_1' =\alpha P$ with $0<\alpha <1$ or for $P_1'=0$. 
This situation is best studied in the original coordinate $\tau$. In the $\tau$ plane.
The equation corresponding to the wall of marginal stability \eq{circlelb1} reduces to a line. It intersects the wall corresponding to the small black hole decays at 
\eq{common-pt} in the 
lower half $\tau$ plane.  It also intersects the real line at 
\begin{equation}
 \tau_1 =  \frac{D}{B} = 
 \frac{ {\rm Im} (Q_1'\cdot w Q\cdot \bar w )} 
{  {\rm Im} ( P_1'\cdot w Q \cdot \bar w +  Q_1' \cdot w P\cdot \bar w)}.
\end{equation}
From the structure of the small black hole decays given in figure 1. it is clear that unless 
this point in the above equation is an integer, it will intersect one the circles corresponding to 
the small black hole decay, in particular the small black hole decay corresponding to the 
bounding circles which join points $(n, 0)$ and $(n+1, 0)$ on the real line. 
Therefore we need  $\frac{D}{B} =n$ to be an integer to 
ensure it does not intersect the bounding small black hole decays.  
When this occurs, a little thought shows that this decay is identical to a small black hole decay  characterized by the matrix
\begin{equation}
\label{matrix-l}
 \left( \begin{array}{cc}
1  & n \\
0 & 1 
\end{array}
\right).
\end{equation}
This is because the line corresponding to the above small black hole decay and the decay 
with $\frac{D}{B} = n$ intersect at the common point \eq{common-pt} and on the real 
axis at $n$. Therefore, both of them must be coincident. 
To summarize we have shown that for $A \neq 0$,  a generic decay does not intersect
the  wall corresponding to the small black hole decay if \eq{class-1} is satisfied. For $A=0$ one
the line corresponding to the generic decay must pass through an integer on the real axis.

{\emph{We  now show  that the class of decays  which satisfy  \eq{class-1} are the 
only physical decays for which the second point  of intersection with the 
small black hole decays  is such that $\tilde \tau_2 \leq 0$. 
That is we prove that the condition   in \eq{class-1} is not only necessary
but also sufficient. }}. 
Note that the decays we discussed when $A=0$ whose walls of marginal stability reduce to 
straight lines also satisfy the condition \eq{class-1} as one can see from \eq{matrix-l} 
Our strategy to prove this will be  by elimination. 
We will show that all the remaining cases 
are such that one can choose $a, b, c, d$ such that $R<0$ or equivalently 
find a small black hole decay which intersects with the generic decay in the 
interior of the upper half plane.  
Without loss of generality we will assume that 
$\frac{p}{q} > \frac{p'}{q'}$ and $q, q'>0$ with $p$, $q$ relatively prime, 
and $p'$, $q'$ relatively prime.

\vspace{.5cm}
\noindent
{\bf{ Case(i)  $r_+ = \frac{p}{q}, r_-  = \frac{p'}{q'}$ with $pq' - p'q = n$}}
\vspace{.5cm}

This case also breaks up into two. Let us first consider the case of $n >{qq'}$
then $r_+ - r_->1$ and there always exists an integer in between $r_+$ and $r_-$.  
From the structure of domains formed by small black hole decays, it can be seen
that there is a  small black hole decay whose wall is a 
straight line passing through every integer on the real axis. 
This line certainly will intersect the circle joining $r_+$ and $r_-$.
Now consider the case for which $n <qq'$, then 
by the main theorem of the linear Diophantine equation discussed in the  appendix \eq{theorem},  
we can find $a, b, c, d  \in \ZZZ $ with $ad -bc =1$ and $d, c>0$ such that 
\begin{eqnarray}
\label{diopeq}
 & & \frac{b}{d} < \frac{p}{q} < \frac{a}{c}, \qquad \hbox{with} \qquad d<q, c<q,  \\ \nonumber
 & & pd - bq =1, \quad  \hbox{and} \qquad aq - pc =1, \qquad ad -bc =1.
\end{eqnarray}
Now consider the points 
\begin{equation}
t_- = \frac{b + m p}{ d+ mq} < \frac{p}{q} < t_+ =  \frac{ a - mp}{c-  mq},
\end{equation}
where $m$ is a positive integer. 
Note that there is a small black hole decay passing through the points $t_-$ and $t_+$
since $ (a -mp)( d+ mq) - ( b+ mp) ( c - mq) =1$.
The distance between $t_-$ and $r_-$ is given by
\begin{equation}
 t_- -r_ - = \frac{ n(d + mq) - q'}{ qq'(d + mq)}.
\end{equation}
To obtain this we have used the equations in \eq{diopeq}. 
By suitably choosing $m$ large enough, it is clear that 
we can ensure
\begin{equation}
 r_- < t_- < r_+ < r_+.
\end{equation}
Now that there is a circle corresponding to small black hole decay joining 
$t_-$ and $t_+$, it is clear 
from the discussion in \eq{int-intersect}
that the circle joining $r_-$ and $r_+$ intersects  the former 
in the interior of the $\tau$ plane. 
We have therefore seen that if $r_+$, $r_-$ are rational such that 
$pq' - p'q =n$ with $n>1$, there is always a small black hole decay intersecting 
the wall passing through $r_+$ and $r_-$ in the interior of the moduli space.

\vspace{.5cm}
\noindent
{\bf{Case(ii)  $r_+$ is irrational} }
\vspace{.5cm}

Let us suppose $r_+$ is irrational.  Then by the  corollary to the Dirichlet's approximation theorem
\eq{corollary} we can find infinite rationals $\frac{p}{q}$ with $p$ and $q$ relatively prime and 
$q>0$ such that 
\begin{equation}
\label{dirichlet}
 \left| r_+ - \frac{p}{q} \right| < \frac{1}{q^2}.
\end{equation}
Now using  the main  theorem on the linear Diophantine equation we know that  there exists
integers $a, b, c, d$ with $ad -bc =1$ satisfying 
\eq{diopeq}.  There is a wall corresponding to a small black hole 
decay passing through $s_- = \frac{b}{d}$ and $s_+ = \frac{a}{c}$. 
Let us consider first the case that
$\frac{p}{q} > r_+$. Then 
\begin{eqnarray}
 r_+ - \frac{b}{d} &=& - \left( \frac{p} {q} - r_+\right)  + \left( \frac{p}{q} - \frac{b}{d} \right), \\ \nonumber
 &>& - \frac{1}{q^2} + \frac{1}{qd} = \frac{q-d}{qd}.
\end{eqnarray}
Where we have used the inequality \eq{dirichlet}. Now since we have $q>d$, we can conclude that 
\begin{equation}
 r_- <  s_- < r_+ < \frac{p}{q} < s_+.
\end{equation}
From \eq{int-intersect} we have,
 the circle corresponding to the small black hole
  decay joining $s_-$ and $s_+$ intersects the circle joining $r_-$ and $r_+$. 
If $\frac{p}{q} <r_+$, then 
\begin{eqnarray}
 \frac{a}{c} - {r_+}  
&=& \left( \frac{a}{c} - \frac{p}{q} \right) + \left( \frac{p}{q} - r_+ \right) , \\ \nonumber
&> & \frac{q - c}{cq}.
\end{eqnarray}
where we have  used the inequality \eq{dirichlet}.
Since $q> c$ we conclude that 
\begin{equation}
 r_- <s_- < \frac{p}{q} < r_+ < \frac{a}{c}.
\end{equation}
Again we have the situation that the circle corresponding to the small black hole decay 
joining $s_-$ and $s_+$ intersects the circle joining 
$r_-$ and $r_+$.
Thus if  $r_+$ is irrational then the wall of marginal stability which
passes through $r_-$ and $r_+$ intersects some wall 
corresponding to a small black hole decay in the interior of the moduli space.

\vspace{.5cm}
\noindent
{\bf{ Case(iii) $r_-$ is irrational}}
\vspace{.5cm}

For this case, the argument to show that there exists a 
wall corrresponding to a small black hole decay  which intersects the 
wall of marginal stability joining $r_-$ and $r_+$ is same as for the Case(ii) discussed earlier.

\vspace{.5cm}
\noindent
{\bf {Case(iv) $r_+$ and $r_-$ are irrational}}
\vspace{.5cm}

It is clear one can again use the argument for either Case(ii) or Case(iii) to show that 
there exists a wall corrresponding to a small black hole decay which intersects the 
wall of marginal stability joining $r_-$ and $r_+$.

We now have exhausted all the possibilities for the values of $r_+$ and $r_-$ and have 
shown that except for the situation when $r_-$ and $r_+$ are rational and 
$pq' -p'q =1$ there is always a small black hole decay which intersects the circle joining
$r_-$ and $r_+$ in the interior of the upper half plane. 
Thus the necessary and sufficient condition that the walls of marginal 
stability never intersect in the interior of the upper half $\tau$ plane 
is when the walls satisfy \eq{class-1}

For completeness
let us now find the point $\tilde \tau_2$ for the class of decays which satisfy
\eq{class-1}. 
We have shown that the second point of intersection of this class of black holes and 
the class of small black holes
 $\left( \begin{array}{cc} a  & b \\ c & d 
                                                            \end{array}\right) $
is determined by \eq{2-intersect}. 
Since $R\geq 0$, the only physically relevant point is when 
$R$ vanishes or $R$ is $\infty$. From the expression of $R$ given in 
\eq{def-ratio} we see that $R$ can vanish at
\begin{eqnarray}
  \frac{p}{q} = \frac{b}{d} \quad \hbox{or} \quad  \frac{p'}{q'} = \frac{b}{d}
\end{eqnarray}
In, this case,  $\tilde \tau_2' =0, \tilde \tau_1' =0$ 
and  the intersection point in the original variables is at 
$\tilde \tau  = \frac{b}{d}$.
$R$ can also be infinity when 
\begin{equation}
 \frac{p}{q} = \frac{a}{c} \quad \hbox{or} \quad \frac{p'}{q'} = \frac{b}{d}.
\end{equation}
In this case  $\tilde \tau'$ is at $i\infty$, while in 
the original $\tau$ variable, the intersection point is at 
$\tau = \frac{a}{c}$. 
Thus we can conclude the wall of marginal stability of 
a small black hole decay intersects with the those 
which satisfy the condition \eq{class-1}  only if the following sets have 
an overlap
\begin{equation}
 \{\frac{a}{c}, \frac{b}{d} \} , \quad \{\frac{p}{q}, \frac{p'}{q'} \}.
\end{equation}
Note that this is the same conditions  obtained by \cite{Sen:2007vb} for the case of small black hole decays 
in ${\cal N}=4$ theories. This is to be expected since we have seen that the 
walls which satisfy \eq{class-1} are coincident with small black hole decays.

\subsection{Walls bounded by walls of small black hole decays}

Using our earlier results, it is now easy to find the conditions on the 
walls so that they are all confined in domain II for figure 1. 
That is the walls are such that they are bounded by the bounding walls 
corresponding to small black hole decays. 
Consider the class of decays with points of intersection on the 
real axis $r_+$ and $r_-$ such that they are 
in the interval $[n, n+1]$ where $n \in \ZZZ$. That is $r_+$ and 
$r_-$ are such that 
\begin{equation}
\label{confine}
n \leq r_+ \leq n+1, \qquad \hbox{and}\qquad n \leq r_- \leq n+1
\end{equation}
Then from the structure of the domains of the small black hole decay
shown in figure 1. , 
we see that such decays never intersect the small black hole decay 
which passes through the points $n$ and $n+1$ in the interior of the
upper half plane.  
This is because  they don't  satisfy the condition \eq{int-intersect} required for
intersection in the interior of moduli space.
Thus they are lie below this small black hole decay and are 
therefore they are confined to domain II. 
Note that
if $r_+ = n+1, r_- = n$, then the decay we are considering coincides
with the small black hole decay. 
Thus all decays  which satisfy \eq{confine} are bounded by small
black hole decays joining the points $(n, 0)$ and $(n+1, 0)$
Therefore if one restricts the moduli and the charges of decay so that 
they satisfy \eq{confine} then the region II in figure 1. is entirely 
free from any decay in this class.

\section{Entropy Enigma decays}

We have seen that 
in these class of ${\cal N}=2$ models the moduli can easily be parametrized in terms
of the complex coordinates $\tau$ and $w$. The 
the BPS mass formula is simple expression in terms of 
charges and these moduli. Furthermore  all walls
of marginal stability are circles or lines in the $\tau$ plane. Therefore it is interesting
to re-examine the phenomenon of  `Entropy Enigma' found 
in \cite{Denef:2007vg,Bates:2003vx, Denef:2007yn} and see how they occur in the $\tau$ plane. 
Briefly the Entropy enigma decays are BPS decays which occur when the 
entropy of the products is parametrically larger than the entropy of the 
parent. We will now  enumerate all the possible charge configurations of the 
decay products which can lead to this 
situation for the class of ${\cal N}=2$ models discussed in this paper.
 Consider the decay
\begin{eqnarray}
\label{enig-decay}
 \left(
\begin{array}{c}
\Lambda Q \\ \Lambda P \end{array} \right)
&=& \left(\begin{array}{c}
         Q_1\\P_1
        \end{array}
\right)
+ \left(\begin{array}{c}
         Q_1\\P_2
        \end{array}
\right), \\ \nonumber
&=& \left(
\begin{array}{c}
 \frac{\Lambda}{2} Q + \Lambda^2 q \\ \frac{\Lambda}{2} P + \Lambda^2 p 
\end{array}
\right)
+ 
\left(
\begin{array}{c}
 \frac{\Lambda}{2} Q - \Lambda^2 q \\ \frac{\Lambda}{2} P - \Lambda^2 p 
\end{array}
\right).
\end{eqnarray}
Since the initial dyon is supersymmetric and a large black hole we have the following
\begin{equation}
 Q^2>0,\quad P^2>0, \quad Q^2P^2 > (Q\cdot P)^2,\quad S= \Lambda^2 \pi \sqrt{ Q^2P^2 - (Q\cdot P)^2},
\end{equation}
where $S$ is the  Hawking-Bekenstein entropy of the black hole. 
To ensure that the 
decay products are  supersymmetric 
and are large black holes in the limit $\Lambda \rightarrow \infty$ 
 we have to impose the  following conditions
$$Q_i^2>0,\qquad  P_i^2>0, \qquad Q_i^2 P_i^2 - (Q_i\cdot P_i)^2>0$$ 
with $i =1, 2$, in the large $\Lambda$ limit.
This leads to the following $3$ cases which exhibit the entropy enigma. 
\begin{enumerate}
 \item \begin{eqnarray}
\label{enig-1}
      & &    q^2>0, \quad p^2>0, \quad q^2p^2 - (q\cdot p)^2 \\ \nonumber
& &  S_{2} = 2\pi \Lambda^4\sqrt{q^2p^2 - (q\cdot p)^2}.
       \end{eqnarray}
Here $S_2$ refers to the sum of the leading entropy of the products in the $\Lambda \rightarrow \infty$
limit.  Note that $q^2>0, p^2>0$ is obtained if one demands $Q_i^2>0, P_i^2 >0$ in the 
$\Lambda\rightarrow \infty$ limit. 
\item 
\begin{eqnarray}
\label{enig-2}
 & & q^2>0, \;  p^2=0,\; p\cdot P =0 \; q\cdot p=0, \; q^2P^2 - (Q\cdot p + q\cdot P)^2  >0 \\ \nonumber
& & S_2 = \Lambda ^2 \pi \sqrt{ q^2P^2 - (Q\cdot p + q\cdot P)^2 }
\end{eqnarray}
Note that for this case,   $p^2=0$ implies that $p\cdot P =0$ on demanding
$P_1^2>0$ and $P_2^2>0$ in the $\Lambda \rightarrow \infty$ limit. 
Furthermore, demanding that that the decay products are large black holes 
(i.e  $Q_i^2 P_i^2 - (Q_i\cdot P_i)^2>0$ ) in the 
$\Lambda \rightarrow \infty$ limit forces $q \cdot p =0$ and $q^2P^2 - (Q\cdot p + q\cdot P)^2  >0$. 
\item
\begin{eqnarray}
\label{enig-3}
& & p^2>0,\;  q^2=0, \; q\cdot Q =0,\;  p\cdot q =0 ,\;  p^2 Q^2 - (Q\cdot p + q\cdot P)^2 \\ \nonumber
& & S_2  = \pi \Lambda^3 \sqrt{ p^2 Q^2 - (Q\cdot p + q\cdot P)^2}
\end{eqnarray}
This is a similar situation to that of case 2. Here $q^2=0$ implies that 
$q\cdot Q=0$ on demanding $Q_1^2 >0$ and $P_2^2>0$ in the $\Lambda\rightarrow\infty$ limit. 
Demanding that the decay products are large black holes in the $\Lambda\rightarrow\infty$ limit
gives rise to the condition $q\cdot p =0$ and $p^2 Q^2 - (Q\cdot p + q\cdot P)^2>0$. 
\end{enumerate}
Finally when one imposes the last possible condition $q^2=0, p^2=0$, then one is forced
to set $q\cdot Q =0, q\cdot P =0, q\cdot p =0, \cdot P$ to ensure that
all decay products are supersymmetric and are large black holes.  Then the leading  entropy of the 
decay products is given by $\pi \lambda^2/2 \sqrt{ Q^2 P^2 - (Q\cdot P)^2}$ which is not 
parametrically larger than the parent black hole. 

We now examine the wall of marginal stability for these decays and show that if 
 the moduli $w$ is generic, that is all the following moduli dependent 
quantities do not scale with $\Lambda$ 
\begin{equation}\
\label{def-generic}
 Q \cdot w, P\cdot w, q\cdot w, p\cdot w \sim O(\Lambda^0),
\end{equation}
then entropy enigma decays are not possible. This phenomenon was observed in 
the specific examples studied in \cite{Denef:2007vg, Denef:2007yn} but was not
shown in general \footnote{See below equation (8) of \cite{Denef:2007yn}}. 
The wall of marginal stability for the decay is determined by the equation \eq{circlelb1} and 
inequality \eq{circlelb2}. On substitution  of the charges of the decay products in \eq{enig-decay}
in these equations  we obtain the following 
\begin{eqnarray}
 \label{enigma-1}
& & \tau\bar\tau  {\rm Im} [ p_\cdot w P \cdot \bar w] - 
\tau_1  {\rm Im} [ (p \cdot w)( Q \cdot \bar w)  +( q\cdot w P\cdot \bar w)]   \\ \nonumber
& & -\tau_2 {\rm Re} [(p \cdot w)( Q \cdot \bar w)  -( q\cdot w P)(\cdot \bar w)] 
+ {\rm Im} [( q \cdot w) ( Q\cdot \bar w)] =0 ,
\end{eqnarray}
and
\begin{equation}
 \label{enigma-2}
\frac{1}{4} |Q\cdot w - \tau P\cdot w|^2  -\Lambda^2 | q\cdot w - \tau p\cdot w|^2 >0.
\end{equation}
As we have seen earlier, the first equation and the equation equation determining the inequality 
are circles. From \eq{intersect} we see that the two circles intersect at points where the central charges of the 
decay products vanish. These points are 
\begin{equation}
\label{enig-intersect}
\tau_+ =  \frac{ (Q + 2\Lambda q)\cdot w}{ (P +  2\Lambda p) \cdot w }, \qquad 
 \tau_-  = \frac{ (Q -  2\Lambda q)\cdot w}{ (P -  2\Lambda p) \cdot w }.
\end{equation}
Note that since both these points correspond to points at which the central charges 
of either of the decay products vanish, these points must lie in the 
lower half $\tau$ plane.

We now analyze  the various cases and show that entropy enigma decays 
at generic values of the $w$ moduli there are no lines or marginal stability 
corresponding to the entropy enigma in the 
limit $\Lambda\rightarrow\infty$.

\vspace{.5cm}
\noindent
{\bf{Generic values of $w$-moduli: ${\rm{ Im}} ( p\cdot w P\cdot \bar w) \neq 0 $}}
\vspace{.5cm}

In the $\Lambda\rightarrow \infty$ limit, 
and at generic values of the $w$ moduli, more specifically 
when the moduli is such that \eq{def-generic} is satisfied,
 we see that the points of intersection of the 
circles coincide \eq{enig-intersect} coincide to $O(\Lambda^{-1} )$ terms. 
\begin{equation}
\label{cent-vani}
 \tau_{\pm}|_{\Lambda \rightarrow \infty}   = \frac{ q\cdot w}{p\cdot w} \left[
 1  \pm\frac{1}{2\Lambda}  \left( \frac{Q\cdot w}{ q\cdot w} \mp \frac{P\cdot w}{ p\cdot w} \right)
+ O(\frac{1}{\Lambda^2} ) \right].
\end{equation}
Therefore  to the leading approximation in $\Lambda$, 
the two circles \eq{enigma-1} and \eq{enigma-2} are tangents to each other.
 Now we need to see if the points on the circle
\eq{enigma-1} satisfy the inequality \eq{enigma-2}
in the large $\Lambda$ limit. 
If at all this wall is physical, it must emerge in the upper half plane and therefore must satisfy the 
condition in \eq{no-intersect}. Let us suppose it does, then at these points 
 $|q\cdot w - \tau \cdot w|$ is of $O(\Lambda ^0)$ and non-zero. The only point it vanishes 
is at \eq{cent-vani} which is in the lower half $\tau$plane. 
If $|q\cdot w - \tau \cdot w|$ is of $O(\Lambda ^0)$ then the second term in 
\eq{enigma-2} is the dominant term in the $\Lambda\rightarrow \infty$ limit and 
it is clear that it never satisfies the inequality \eq{enigma-2}. Therefore 
\eq{enigma-1} can never be a physical wall of marginal stability for generic 
values of moduli which satisfy \eq{def-generic} in the $\Lambda\rightarrow\infty$ limit.

\vspace{.5cm}
\noindent
{\bf{Generic values of $w$-moduli: ${\rm{ Im} }( p\cdot w P\cdot \bar w) =0 $}}
\vspace{.5cm}

In general the condition ${\rm{ Im}} ( p\cdot w P\cdot\bar w) =0 $
imposes an additional condition on the moduli space, and therefore the 
corresponding wall of marginal stability will not be a co-dimension one surface in the 
moduli space. Therefore the only way ${\rm{ Im}} ( p\cdot w P\cdot \bar w) $ can vanish is 
$p= \alpha P, \alpha \neq 0$. 
 Then the circle \eq{enigma-1} reduces to the straight line  given by 
\begin{eqnarray}
 \label{enig-line}
& & - \tau_1  {\rm Im} [ n (P \cdot w  Q \cdot \bar w)  +( q\cdot w P \cdot \bar w)] 
 -\tau_2 {\rm Re} [n (P \cdot w  Q \cdot \bar w)  -( q\cdot w P)(\cdot \bar w)]  \nonumber \\
& & + {\rm Im} [ q \cdot w  Q\cdot \bar w] =0 ,
\end{eqnarray}
while the inequality in \eq{enigma-2} reduces to 
\begin{equation}
 \label{enigma-ineq}
\frac{1}{4} |Q\cdot w - \tau P\cdot w|^2  -\Lambda^2 | q\cdot w - n \tau P\cdot w|^2 >0.
\end{equation}
Again the line  in  \eq{enig-line} and  the circle determining the inequality in 
\eq{enigma-ineq} intersect each other at the point where the central charges vanish.
This is given by
\begin{eqnarray}
\label{pm-enig}
 \tau_\pm &=& \frac{ (Q \pm 2\Lambda q)\cdot w}{ P\cdot w (1 \pm  2\alpha \Lambda )  },
\\ \nonumber
&=& \frac{ q\cdot w}{\alpha  P \cdot w} + O(\frac{1}{\Lambda}  ) .
\end{eqnarray}
These points must lie in the lower half $\tau$ plane even in the $\Lambda\rightarrow\infty$ limit 
\footnote{Note that  
$\tau_\pm =( Q \pm 2\Lambda q)\cdot w/ P\cdot w $ for $\alpha =0$.
Thus in $\Lambda \rightarrow \infty$ limit it is  impossible to ensure that 
both these points lie in the lower half
$\tau$ plane for generic values of $w$. Therefore for generic values of $w$ such decays 
are not BPS}.
For generic values  of 
the moduli $w$, that is when \eq{def-generic} is satisfied,  
the part of the line \eq{enig-line} which emerges in the upper half plane is such that
$| q\cdot w - n \tau P\cdot w| \sim O(\Lambda^0)$. This is because this terms vanishes only
at $\tau_{\pm}$ in the $\Lambda\rightarrow \infty$ limit and these point
lie in the lower half $\tau$ plane. 
This implies the second term in \eq{enigma-ineq} is always dominant in the putative wall of 
marginal stability \eq{enig-line} which emerges in the upper half plane. 
Thus the inequality \eq{enigma-ineq} can never be satisfied and \eq{enig-line} does not
represent a physical wall of marginal stability at generic values of moduli $w$.

\section{Conclusions}
We have studied various properties of walls of marginal stability in 
 ${\cal N}=2$ models with the moduli space given by \eq{coset}.
To study these properties we have used the mass formula for BPS states 
obtained from the classical moduli space of these theories.  
The list of the properties we have found are listed  in the introduction. 
Using these properties we have isolated a class of decays with walls which always lie 
in the region bounded by small black hole decays. These walls always lie in 
region II of figure \ref{fig-domains}.  We hope these properties will be useful in constructing and also testing 
possible proposals for BPS spectrum in these models. 

We showed that these models do not admit entropy enigma decays for generic
values of moduli which satisfy \eq{def-generic}. It will be interesting to find 
explicit examples of entropy enigma decays by appropriate scaling of the $w$ moduli 
by the parameter $\Lambda$. Such examples can complement the
examples found in \cite{Denef:2007vg,Denef:2007yn}. 

\acknowledgments
The author thanks Ashoke Sen for insightful discussions at various stages 
of this project and also for comments on the manuscript.
 He thanks Apoorva  Patel for a discussion on 
estimates of real numbers by rationals and providing a useful reference
on this topic.

\appendix
\section{Some number theory}

In this appendix we recall some theorems from number theory which are
 used in this paper

\subsection{Dirichlet's approximation theorem}

{\emph {For each $\alpha$ belonging to the real and $N$ a positive integer, there are $n \in \ZZZ$,  $(n \leq N)$ 
and $p\in \ZZZ$ such that }} 
\begin{equation}
 \left| \alpha - \frac{p}{n} \right| < \frac{1}{Nn}, \quad \hbox{i.e.} \quad
| n\alpha - p | < \frac{1}{N}
\end{equation}

\vspace{.5cm}
\noindent
{\bf Corollary to Dirichlet's approximation theorem}
\vspace{.5cm}

\noindent
{\emph{If $\alpha$ is irrational, then there are infinitely many rationals 
(with strictly increasing denominators)  $\frac{p}{n}$, $n>0$ with 
$p$ and $n$ relatively prime such that  } }
\begin{equation}
\label{corollary}
 \left| \alpha - \frac{p}{n} \right| < \frac{1}{n^2}
\end{equation}

\noindent
The number $\frac{p}{n}$ is called the D-approximation to $\alpha$.
In fact as an  another  corollary one can also  show that if $\alpha$ is rational, it has only 
a finite number of D-approximations.
For proof of the Dirichlet approximation theorem and its corollaries 
see  \cite{Apostol}, see also \cite{Kim} for a short review.

\subsection{Main theorem on the linear Diophantine equation}

{\emph {For each each rational of the form  $\frac{p}{q}$  $q>0$ and $p$, $q$ relatively prime, 
there are $a, c$ such that $aq - cp =1$.  }}

We reproduce the proof of this  theorem from \cite{Kim} with a small modification necessary for our purpose.
Let $\alpha = \frac{p}{q}$. 
If $q=1$, there $a - cp$ is solved by setting $c =0$ and $a =1$. 
Therefore without loss of generality we may assume that $q\geq 2$. 
Applying the Dirichlet approximation theorem, with $N = q-1$, there are
$c \in \ZZZ$ and $c>0, c\leq N = q-1 $ and $a \in \ZZZ$ such that 
\begin{equation}
| \alpha c - a| = \left|\frac{p}{q} c  - a \right| < \frac{1}{N} = \frac{1}{q-1},
\end{equation}
multiplying it by $q$ leads to 
\begin{equation}
 | pc -aq | < \frac{q}{ q-1} = 1 + \frac{1}{q-1} \leq 2.
\end{equation}
Since $pc -aq \in \ZZZ$, this implies that $pc -aq | \leq 1$.  The case $pc -aq =0$ is excluded.
Because, this implies that 
\begin{equation}
 \alpha = \frac{p}{q} = \frac{a}{c} .
\end{equation}
and $c\leq N = q-1\leq q$. This contradicts the assumption that $p$ and $q$ are 
relatively prime.  
Thus the only possibility is 
$ aq - pc = \pm 1 $.
Let us consider the  case
\begin{equation}
\label{diop1}
a q -pc = 1, 
\end{equation}
then we already have the proof of the theorem. In addition to this  let us define the integers
\begin{equation}
 b = p -a, \qquad  d =  q -c.
\end{equation}
Note that  $q>d>0$ since $q>c>0$. Then from \eq{diop1}  we see that
\begin{equation}
\label{diop2}
 bq - p d =1, \qquad ad -bc =1.
\end{equation}
We therefore have shown the existence of the ratios $\frac{b}{d}$ and 
$\frac{a}{c}$   in the following   order
\begin{equation}
 \frac{b}{d}<  \frac{p}{q} < \frac{a}{c}
\end{equation}
which satisfies \eq{diop1} and \eq{diop2}. 
Let us now consider the case 
\begin{equation}
 \label{diop3}
pc -aq =1.
\end{equation}
Then we again define the integers
\begin{equation}
 b = p-a, \qquad d = q-c,
\end{equation}
again $d>0$ since $c<q$.  From \eq{diop4} we see that
\begin{equation}
\label{diop4}
 bq - p d =1.
\end{equation}
Thus now we see that the integer $b$ plays the role of $a$  and $d$ plays the role of $c$ required by the
theorem. In addition we also have
\begin{equation}
\label{diop5}
 bc -ad =1.
\end{equation}
which follows from \eq{diop3}.
Thus in this case we have the ratios in the following increasing order
\begin{equation}
 \frac{a}{c}< \frac{p}{q} < \frac{b}{d},
\end{equation}
satisfying \eq{diop3}, \eq{diop4}, \eq{diop5}

{\it We have  proved the theorem and also  shown that given the ratio $\frac{p}{q}$, $q>0$ and 
$p, q$ relatively prime, there exists two ratios
one lesser and one greater than $\frac{p}{q}$ say $\frac{b}{d}$ and $\frac{a}{c}$ respectively 
with $q>d>0, q>c>0$ such that }
\begin{equation}
\label{theorem}
 ad -bc =1, \qquad pd - bq =1, \qquad aq - pc =1. 
\end{equation}

\bibliography{mstable}
\bibliographystyle{JHEP}
\end{document}